\documentclass[a4paper,prd,twocolumn,showpacs,superscriptaddress,floatfix,nofootinbib]{revtex4-2}

\usepackage{preamble}
\usepackage{orcidlink}
\usepackage{lipsum} \usepackage{graphicx} \usepackage{epsf} \usepackage{epsfig}
\usepackage{amssymb,amsmath,amsfonts} 
\usepackage{amssymb} \usepackage{times} \usepackage{bm}
\usepackage{pdfpages}
\makeatletter
\AtBeginDocument{\let\LS@rot\@undefined}
\makeatother

\usepackage{hyperref} \hypersetup{
colorlinks=true,        
linkcolor=blue,         
citecolor=cyan,         
}

\usepackage{graphicx}  
\usepackage{dcolumn}   
\usepackage{bm}        
\usepackage{amsfonts,amsmath,amssymb,mathrsfs}
\usepackage{color}
\usepackage{natbib,times}
\usepackage{mathtools}
\usepackage{xspace}

\newcommand{\fuka}{\texttt{FUKA}\xspace}
\newcommand{\kadath}{\texttt{KADATH}\xspace}

\newcommand{\cf}{cf.,~}
\newcommand{\ie}{i.e.,~}
\newcommand{\eg}{e.g.,~}

\begin{document}
 \title{Black hole--neutron star binaries with high spins and large mass asymmetries: \\
 	I. Properties of quasi-equilibrium sequences}
 \author{Konrad Topolski\, \orcidlink{0000-0001-9972-7143}}
\affiliation{Institut f\"ur Theoretische Physik, Goethe Universit\"at,
Max-von-Laue-Str. 1, 60438 Frankfurt am Main, Germany}

\author{Samuel D. Tootle\, \orcidlink{0000-0001-9781-0496}}
\affiliation{Department of Physics, University of Idaho, Moscow, ID 83843, USA}
\affiliation{Institut f\"ur Theoretische Physik, Goethe Universit\"at,
Max-von-Laue-Str. 1, 60438 Frankfurt am Main, Germany}

\author{Luciano Rezzolla\, \orcidlink{0000-0002-1330-7103}}
\affiliation{Institut f\"ur Theoretische Physik, Goethe Universit\"at,
Max-von-Laue-Str. 1, 60438 Frankfurt am Main, Germany}
\affiliation{Frankfurt Institute for Advanced Studies, Ruth-Moufang-Str. 1,
60438 Frankfurt, Germany}
\affiliation{School of Mathematics, Trinity College, Dublin 2, Ireland}

\date{\today}

\begin{abstract}
Black hole -- neutron star (BHNS) mergers are a promising target of
current gravitational-wave (GW) and electromagnetic (EM) searches, being
the putative origin of ultra-relativistic jets, gamma-ray emission, and
r-process nucleosynthesis. However, the possibility of any EM emission
accompanying a GW detection crucially depends on the amount of baryonic
mass left after the coalescence, \ie whether the neutron star (NS)
undergoes a ``tidal disruption'' or ``plunges'' into the black hole (BH)
while remaining essentially intact. As the first of a series of two
papers, we here report the most systematic investigation to date of
quasi-equilibrium sequences of initial data across a range of stellar
compactnesses $\mathcal{C}$, mass ratios $q$, BH spins $\chi_{_{\rm
    BH}}$, and equations of state satisfying all present observational
constraints. Using an improved version of the elliptic initial-data
solver \fuka, we have computed more than $1000$ individual configurations
and estimated the onset of mass-shedding or the crossing of the innermost
stable circular orbit in terms of the corresponding characteristic
orbital angular velocities $\Omega_{_{\rm MS}}$ and $\Omega_{_{\rm
    ISCO}}$ as a function of $\mathcal{C}, q$, and $\chi_{_{\rm BH}}$. To
the best of our knowledge, this is the first time that the dependence of
these frequencies on the BH spin is investigated. In turn, by setting
$\Omega_{_{\rm MS}} = \Omega_{_{\rm ISCO}}$ it is possible to determine
the separatrix between the ``tidal disruption'' or ``plunge'' scenarios
as a function of the fundamental parameters of these systems, namely, $q,
\mathcal{C}$, and $\chi_{_{\rm BH}}$. Finally, we present a novel
analysis of quantities related to the tidal forces in the initial data
and discuss their dependence on spin and separation.
\end{abstract}
\maketitle

\section{Introduction}
\label{sec:introduction}

There is now convincing evidence obtained via gravitational-wave (GW)
detections that black hole -- neutron star (BHNS) binary systems exist in
nature and undergo merger events~\cite{Abbott2021}.  More so,
GW events continue to provide evidence for binaries in nature to include
BHs in the range of $3-5\, M_{\odot}$~\cite{LIGOScientific2024}, a
``mass gap'' that is clearly being filled by continued GW detections. As in the
case of neutron star--neutron star (NSNS) binary systems, such as
GW170817~\cite{LIGOScientific:2017vwq}, a rich set of electromagnetic
(EM) counterparts are expected from a BHNS merger~\cite{Colombo2023}.

In light of the difficulties of differentiating between a black hole --
black hole (BBH) and a BHNS system based on the GW signal alone, binaries
in which the neutron star (NS) undergoes a tidal disruption are of
particular interest. The process of tidal disruption in numerical
simulations has been shown to leave behind a massive torus
\cite{Etienne2007b, ShibataTaniguchi2008, Etienne:2008re, Kyutoku2010,
  Kyutoku2011, Foucart2011, Kawaguchi2015, Foucart2019, Most2020e}, whose nontrivial
interactions with the magnetic field~\cite{Etienne2012} and the presence
of neutrino radiation~\cite{Kyutoku2018} potentially drive a variety of
discernible EM signals~\cite{Kyutoku2013} on a wide range of timescales,
ranging from hundreds of milliseconds to weeks (see, for example
\cite{Barbieri2019}). Understanding the conditions under which a tidal
disruption occurs, quantifying its effects on the GW signal and its
feedback on the equation of state (EOS)~\cite{Pannarale2011, Lackey2012,
  Lackey2013}, predicting the mass of the remnant black hole (BH) and its
spin~\cite{Pannarale2012, Pannarale2013b, Zappa2019}, as well as the mass
of the formed accretion disc and dynamical ejecta~\cite{Pannarale2010,
  Foucart2012, Kyutoku2015, Foucart2018b, Kruger2020}, have all been
studied in numerous papers in the context of general-relativistic
magnetohydrodynamical (GRMHD) simulations.
Furthermore, a number of investigations have looked into fully three-dimensional
evolution of the post-merger BH and torus configuration on timescales of
several hundred milliseconds~\cite{Rezzolla:2010, Paschalidis2014,
  Ruiz2018, Most2021a} to several seconds~\cite{Hayashi2022,
  Gottlieb2023a, Hayashi2023}, predicting substantial Poynting fluxes
owing to the Blandford-Znajek energy extraction mechanism
\cite{Blandford1977}, coincident with a collimated outflow with
appreciable Lorentz factors (albeit limited due to numerical difficulties
associated with the conservative-to-primitive recovery and insufficiently
low atmosphere levels) and high magnetization~\cite{Gottlieb2023a} far
away from the jet launching site (see~\cite{Duez2024} for a summary of
the most recent developments with regards to seconds-long evolutions in
the post-merger phase).

This work is the first in a series of two papers in which we investigate the
properties of BHNS systems under a variety of gravitational and
microphysical conditions. This paper (paper I), in particular, reports
the study of quasi-equilibrium (QE) sequences of BHNS binaries obtained
as solutions to the constraint equations on an initial hypersurface under
an assumption of time-symmetry and orbital circularity. Studies of this
type have a long history beginning with the seminal work
by Taniguchi and collaborators~\cite{Taniguchi2006,
  Taniguchi07, Taniguchi:2008a}, who constructed QE sequences using the BH
excision approach, for a simple polytropic EOS with the adiabatic index
$\Gamma=2$~\cite{Rezzolla_book:2013}, and provided estimates for the
orbital angular velocity at tidal disruption and innermost stable
circular orbit. At the same time, Grandclement~\cite{Grandclement06}
pursued the same goal and obtained similar results for the binding
energy, with some differences in the quantity predicting the tidal
disruption, while Kyutoku and collaborators~\cite{Kyutoku2009}
investigated sequences where the BH is represented by a puncture and
obtained similar results for the mass-shedding diagnostics, but noted
deviations in the binding energy when compared to the excision approach.

The extension of these initial studies to more realistic configurations
which include both a spinning primary (\ie the BH) and realistic EOSs has
been relatively limited and aimed mostly at showcasing the abilities and
convergence properties of new and advanced initial-data
solvers~\cite{Henriksson2014, Tacik:2016zal, Rashti2021, Papenfort2021b}.
To bridge this gap, we here report the most systematic investigation to
date of QE sequences of initial data across a range of
stellar compactnesses $\mathcal{C}:=M/R$, with $M$ and $R$ being the
stellar gravitational mass and radius, respectively; binary mass ratios $q:=M_{_{\rm
    NS}}/M_{_{\rm BH}}$, with $M_{_{\rm BH}}$ and $M_{_{\rm NS}}$ being
the BH and NS gravitational masses, respectively; (dimensionless) BH spins $\chi_{_{\rm
    BH}}$, and EOSs satisfying all present observational constraints. For
simplicity, and to ensure a reasonable size of the space of parameters, we
will not investigate QE sequences with spinning NSs
and set $\chi_{_{\rm NS}}=0$ accordingly. A follow-up study will also
discuss the influence of the secondary spin.

This paper I has a number of different purposes. First, it highlights the
capabilities of an improved version of the publicly available
Frankfurt-University-Kadath (\fuka)~\cite{Papenfort2021b,Tootle2024a} codes,
which has been shown to yield accurate initial data under a number of
conditions in the binary~\cite{Tootle2021, Papenfort:2022ywx} and that is
becoming a reference choice in many numerical-relativity
groups~\cite{Kuan2023, Kuan2023a, Markin2023, Rosswog2023, Chen2024,
  Izquierdo2024}. Second, it offers a comprehensive view of the
dependence of the onset of mass-shedding or the crossing of the innermost
stable circular orbit in terms of the corresponding characteristic
orbital angular velocities $\Omega_{_{\rm MS}}$ and $\Omega_{_{\rm
    ISCO}}$ as a function of $\mathcal{C}, q$, and $\chi_{_{\rm BH}}$. In
this way, and by setting $\Omega_{_{\rm MS}} = \Omega_{_{\rm ISCO}}$, it
is possible to determine the separatrix between the ``tidal disruption''
and ``plunge'' scenarios as a function of $q$, $\mathcal{C}$, and
$\chi_{_{\rm BH}}$. Finally, the paper guides the set of initial conditions
for the dynamical evolution of a large number of BHNS systems that will
be presented in paper II~\cite{Topolski2024c}.

The paper is organized as follows. In Sec.~\ref{sec:initial_data}, we
briefly summarise the elliptic equations which are solved to obtain the
QE sequences and define the key diagnostic quantities that
allow us to judge the quality of the solutions. In
Sec.~\ref{sec:methods_and_parameter_space}, we shortly describe the
numerical setup in the \fuka code, outline the covered space of
binary parameters, and the methods used to extract the relevant
information from a sequence. Therein, a novel analysis of tidal forces on
an initial data hypersurface is also presented. The results for
irrotational sequences for three cold EOSs are presented in
Sec~\ref{sec:results}, preceded by a comparison of $\Gamma=2$ polytrope
predictions by~\cite{Taniguchi:2008a}. The section concludes with a study
of QE sequences with a spinning BH and includes new results on the spin
dependence of the mass-shedding diagnostic and the location of an
effective innermost stable circular orbit. Also discussed is the
nontrivial dependence of binding energy values on BH spin for realistic
mass-ratio sequences. We summarise the main findings of our work in
Sec.~\ref{sec:summary} and suggest possible improvements and avenues that
could be explored. Throughout the paper, we use the geometrical units $G
= c = 1$, where $G$ and $c$ are the gravitational constant and the speed
of light, respectively. Greek (Latin) indices run from $0$ ($1$) to $3$.

\section{Mathematical setup: the XCTS system in quasi-equilibrium}
\label{sec:initial_data}

The foundation of this work is the use of the publicly available \fuka
initial data code originally described in~\cite{Papenfort2021} with
significant advancements included in the second version of
\fuka~\cite{Tootle2024a}, which provides a highly robust solver
infrastructure to compute incredibly challenging initial data sequences
such as those which will be discussed in Sec~\ref{sec:results}, in
addition to BBH and BNS configurations. Here we will briefly describe the
system of equations solved by \fuka followed by an overview of the
numerical implementation in Sec.~\ref{sec:methods_and_parameter_space}.

To solve the initial value problem on a spacelike hypersurface in GR,
\fuka computes solutions to the widely used eXtended Conformal Thin
Sandwich (XCTS) formalism~\cite{Pfeiffer:2002iy, York99}, which is
customarily applied for generating QE configurations, \ie quasi-circular
orbits and approximately time-symmetric. For a spacetime to be in
\textit{quasi-equilibrium}, the existence of a helical symmetry vector
$\xi^{\mu} := \partial_{t}^{\mu} + \Omega \partial_{\varphi}^{\mu}$ is
assumed, where the (inertial) coordinate time vector $\partial_{t}^{\mu}
:=\alpha n^{\mu} + \beta^{\mu}$ is built from the lapse function
$\alpha$, the normal vector to the hypersurface $n^{\mu}$, and the shift
vector $\beta^{\mu}$. Similarly, the azimuthal component in
$\boldsymbol{\xi}$ is expressed in terms of the orbital angular velocity
of the binary $\Omega$, with $\partial_{\varphi}$ the vector field
generating the rotational symmetry. Should such a vector field
$\boldsymbol{\xi}$ exist, all tensor fields $\boldsymbol{\mathcal{T}}$ in
the spacetime would be invariant under the symmetry group it generates,
which is expressed as the vanishing of Lie derivatives
$\mathcal{L}_{\boldsymbol{\xi}} \boldsymbol{\mathcal{T}} = 0$. The
existence of a global helical symmetry implies in particular that
$\boldsymbol{\xi}$ is a Killing vector field. We will henceforth often
refer to $\boldsymbol{\xi}$ as a Killing vector, understanding that it
can represent a more general and deeper symmetry.

In this work we consider only conformally flat initial data where the
purely spatial metric, $\gamma_{ij}$, is approximated as $\gamma_{ij} =
\psi^{4} \tilde{\gamma}_{ij}$. This decomposition drastically simplifies
the system of equations as $\tilde{\gamma}_{ij}$ is a fixed conformal
metric with a flat connection ($\rm{Riem}[\tilde{\gamma}]=0$, \eg
$\delta_{ij}$) and $\psi$ is the conformal factor that simply rescales
$\tilde{\gamma}_{ij}$ locally. The extrinsic curvature tensor $K_{ij}$
is correspondingly decomposed into the conformal trace-free part
$\hat{A}_{ij}$ and the trace $K:= \gamma^{ij} K_{ij}$ as
\begin{equation}
K_{ij} = \psi^{-2} \hat{A}_{ij} + \frac{1}{3} K \gamma_{ij}\,.
\end{equation}
In this way, the condition of QE simplifies the system of
equations considerably and is imposed by demanding that the derivatives
of the trace of the extrinsic curvature and the conformal metric vanish,
\ie $\mathcal{L}_{{\boldsymbol{\xi}}} K = 0 =
\mathcal{L}_{\boldsymbol{\xi}} \tilde{\gamma}_{ij}$. Such a specification,
combined with enforcing the maximal slicing condition, $K=0$, completes the
choice of the \textit{free data} of the XCTS scheme.

The elliptic equations of the initial-data system that are implemented in
\fuka provide a solution for the conformal factor $\psi$, the lapse $\alpha$, the
inertial shift vector $\beta^{i}$, the extrinsic curvature tensor
$\hat{A}_{ij}$, the corotating fluid velocity $V^{i}$ and the logarithm
of the specific enthalpy $H := \ln h$. The set of constraint equations
$\{ \mathcal{H},\mathcal{C}_{\alpha\psi}, \mathcal{M}^{i},
\mathcal{C}_{\rm v. p.} \}$ to be computed simplifies to\footnote{Note
that in practice the solver uses the equations multiplied by the ratio of
pressure to density $p / \rho$ in order to improve the performance of
spectral methods in the presence of matter fields.}
\begin{align}
  \label{eq:XCTS_constraints_1}
  \mathcal{H} &\coloneqq \tilde{D}^{i}\tilde{D}_{i}\psi +
  \frac{1}{8}\psi^{-7}\hat{A}^{ij}\hat{A}_{ij} + 2\pi \psi^{5}
  E\,,\\
  \label{eq:XCTS_constraints_2}
  \mathcal{C}_{\alpha\psi} &\coloneqq
  \tilde{D}^{i}\tilde{D}_{i}(\alpha\psi)
  -\frac{7}{8}\alpha\psi^{-7}\hat{A}^{ij}\hat{A}_{ij} - 2\pi \alpha
  \psi^{5} (E + 2S)\,,\\
  \label{eq:XCTS_constraints_3}
  \mathcal{M}^{i} &\coloneqq
  \tilde{D}^{j}\tilde{D}_{j}\beta^{i} +
  \frac{1}{3}\tilde{D}^{i}\tilde{D}_{j}\beta^{j} -
  2\hat{A}^{ij}\tilde{D}_{j}(\alpha \psi^{-6}) \\
  \label{eq:XCTS_constraints_4}
  &\phantom{\coloneqq}
  \nonumber -16\pi \alpha \psi^{4} p^{i}\,,\\
  \mathcal{C}_{\rm v. p.}
  &\coloneqq \psi^{6} W V^i \tilde{D}_i H + \frac{dH}{d\ln\rho}
  \tilde{D}_{i}(\psi^{6} W V^i)\,,
\end{align}
where the matter source terms $(E, S, p^{i})$ are the same as
in~\cite{Papenfort2021b} and read
\begin{align}
	\label{eq:Matter_source_1}
	E &\coloneqq \rho h W^{2} - p\,,\\
	\label{eq:Matter_source_2}
	S &\coloneqq 3p + (E+p)U^{2}\,, \\
	\label{eq:Matter_source_3}
	p^{i}&\coloneqq \rho h W^{2} U^{i}\,,
\end{align}
and
\begin{align}
  \hat{A}_{ij} &:= \frac{\psi^6}{2 \alpha} (\tilde{D}^i \beta^j + \tilde{D}^j
      \beta^i - \frac{2}{3} \tilde{\gamma}^{ij} \tilde{D}_k \beta^k)\,. &&
      \label{equ:SYS_Aij}
\end{align}

The relativistic specific enthalpy is $h\coloneqq 1 + \epsilon + p/\rho$
with pressure $p$, density $\rho$ and specific internal energy $\epsilon
$. In the above, $U^{i}$ is the spatial projection of the four-velocity
of the fluid, $W\coloneqq (1-U^2)^{-1/2}$ is the associated Lorentz
factor, and $\tilde{D}^{i}$ denotes the covariant derivative compatible
with the conformal metric $\boldsymbol{\tilde{\gamma}}$. We have also
defined the spatial corotating fluid velocity as $V^{i} = \alpha U^{i} -
\xi^{i}$. Equation~\eqref{eq:XCTS_constraints_4} determines the velocity
potential (v.p.) and derives from the continuity equation, whereas
$\mathcal{C}_{\alpha\psi}$ is obtained directly from the dynamical
evolution equation for the trace of the extrinsic curvature by setting
the right-hand side of $\mathcal{L}_{\boldsymbol{\xi}}K$ and $K$ itself
to zero. \fuka solves the elliptic
equations~\eqref{eq:XCTS_constraints_1}--\eqref{eq:XCTS_constraints_4} to
a desired precision and after adequate boundary conditions are imposed
(see~\cite{Papenfort2021b} for more details). Additionally, the
approximate first integral of the relativistic Euler equation is imposed
onto the NS matter
\begin{align}
 \log(h \alpha W^{-1}+ \tilde{D}_i \Phi V^i) = C\,,
\end{align}
where the constant $C$ is fixed depending on the value of the specific
enthalpy $h$ at the centre of the star. Finally, we must provide an EOS
to close the system of equations. \fuka includes an EOS module that
supports the use of (piecewise-)polytropic EOSs as well as 1D tabulated
EOSs for realistic models which is yet another critical component that
has enabled the novel exploration of this work.

The assumed stationarity of the solution can be validated by inspecting the
violation of $\boldsymbol{\xi}$ to represent an asymptotically timelike
Killing vector. More specifically, we compute the \textit{virial
  error}, defined as
\begin{align}
  \delta M := 1 - \frac{M_{_{\rm K}}}{M_{_{\rm ADM}}}\,,
  \label{eq:virial_error}
\end{align}
where $M_{_{\rm ADM}}$ and $M_{_{\rm K}}$ are the ADM and Komar masses,
respectively. Under the assumption of conformal flatness
these masses are defined as~\cite{Papenfort2021b}
\begin{align}
  &M_{_{\rm ADM}} \coloneqq -\frac{1}{2\pi}\int_{S_{\infty}} D_{i}\psi \;s^i d^{2}\sigma\,,
  \label{eq:adm_mass}\\
  &M_{_{\rm K}} \coloneqq \frac{1}{4\pi}\int_{S_{\infty}} D_{i}\alpha \; s^i d^{2}\sigma\,,
  \label{eq:komar_mass}
\end{align}
where $s^i$ is the spatial unit vector normal to the sphere of integration.
We note that \fuka is well suited for computing the virial error as the
physical space is decomposed in a multi-grid scheme with the outer
spherical domain utilizing compactified spatial coordinates such that the
outer spherical boundary, $S_{\infty}$, is mapped to spatial infinity
rather than an arbitrary large number (see~\cite{Grandclement09} for
details).

\begin{figure*}
  \center
  \hspace{-1.0cm}
  \includegraphics[scale=0.75]{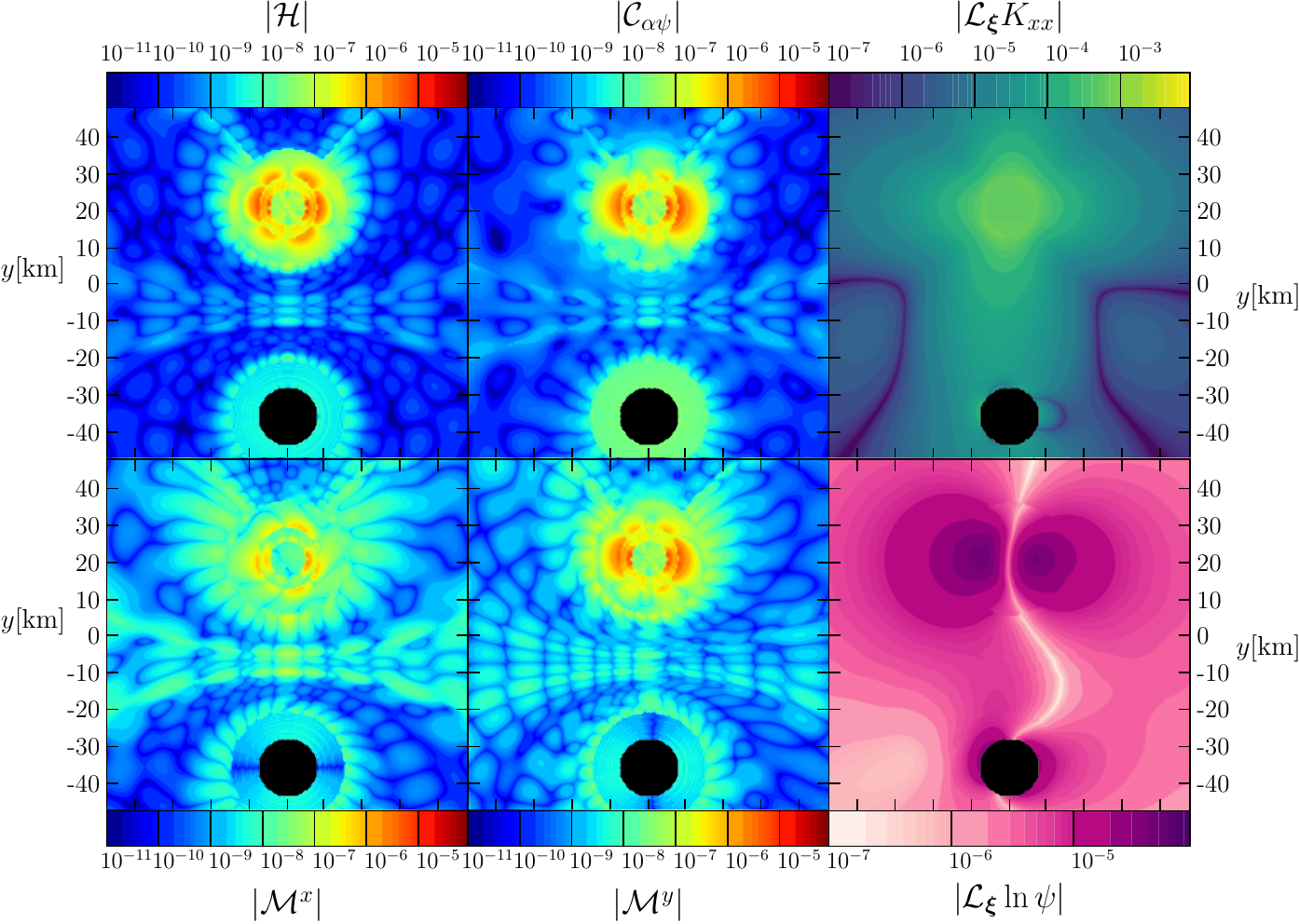}
	\caption{The representative monitoring quantities $\mathcal{H}$
          (top-left panel), $\mathcal{M}^x$ (bottom-left panel),
          $\mathcal{C}_{\alpha\psi}$ (top-middle panel), $\mathcal{M}^y$
          (bottom-middle panel) relative to constraints
          Eqs.~\eqref{eq:XCTS_constraints_1} --
          \eqref{eq:XCTS_constraints_4} of the XCTS system. The data
          refers to a typical BHNS configuration at a separation of
          $d=56.865\,{\rm km}$, with a total mass of $7.800\,M_{\odot}$,
          where $M_{_{\rm NS}}=1.300\,M_{\odot}$, mass ratio $Q :=
          q^{-1}=5$ and where the NS matter is modelled by the
          intermediate EOS. Shown in the top and bottom panels on the
          right are two novel monitoring quantities measuring the
          deviation of $\boldsymbol{\xi}$ from being a true helical Killing
          vector field. In all cases, black circles denote the apparent
          horizon of the BH.}
  \label{fig:BHNS_XCTS_constraints_set}
  \end{figure*}

As pointed out in Ref.~\cite{Foucart2008}, there exists a tension between
the QE assumption for the fluid and the choice $K=0$ in the XCTS system
of equations. This tension follows from deriving the continuity equation
(\cf Eqs.~(50) and (53) in~\cite{Teukolsky98}), where, among others, the
term $\mathcal{L}_{\boldsymbol{\xi}}\ln \psi$ is set to
zero\footnote{Under the conformal decomposition, the term
$\mathcal{L}_{\boldsymbol{\xi}}\ln \psi$ contributes to the last term in
Eq.~(213) in Ref.~\cite{Tichy2017}.}, which is not strictly enforced as
part of the constraint equations despite it also being necessary for
``true'' QE. Such a quantity also appears in the discussion regarding the discrepancy
between QE boundary conditions for binary BHs and spacetime stationarity
in Ref.~\cite{Cook:2004kt}. Its enforcement has been found to cause
non-invertibility of the differential operator associated
with the momentum constraint or lead to instabilities
otherwise~\cite{Pfeiffer:thesis, Cook:2004kt}.
We calculate this term using the identity (see, \eg Eq.~(6.70) in
Ref.~\cite{Gourgoulhon2012})
\begin{align}
  \label{eq:evol_eq_conf_factor}
  \mathcal{L}_{\boldsymbol{\xi}}\ln \psi = \frac{1}{6}\big{(} \tilde{D}_{i}B^{i} -
  \alpha K\big{)} + \mathcal{L}_{\boldsymbol{B}}\ln \psi\,,
\end{align}
where $B^{i} :=\beta^{i} + \Omega \partial_{\varphi}^{i}$ is the
so-called corotational shift; note that the last term in
\eqref{eq:evol_eq_conf_factor} appears when the divergence
of the shift vector is taken with respect to the conformal spatial metric $\tilde{\gamma}_{ij}$
as opposed to the spatial metric $\gamma_{ij}$ as in
Ref.~\cite{Foucart2008}. The same equation as here is also reported, \eg in
Ref.~\cite{Pfeiffer:thesis,Cook:2004kt}. The left-hand side of
Eq.~\eqref{eq:evol_eq_conf_factor} can be seen as a residual of the
equilibrium condition $\mathcal{L}_{\boldsymbol{\xi}} \gamma_{ij}=0$,
which can be written more explicitly as
\begin{align}
 0 = \mathcal{L}_{\boldsymbol{\xi}} \gamma_{ij} = 4 \psi^{4}
 (\mathcal{L}_{\boldsymbol{\xi}} \ln \psi) \tilde{\gamma}_{ij} + \psi^{4}
 \mathcal{L}_{\boldsymbol{\xi}}\tilde{\gamma}_{ij}\,.
\end{align}
Clearly, the choice of free data $\mathcal{L}_{\boldsymbol{\xi}}
\tilde{\gamma}_{ij}=0$ will only remove the second, but not the first
term on the right-hand side. Therefore, $\mathcal{L}_{\boldsymbol{\xi}} \ln
\psi$ provides a simple and yet useful estimate of the deviation from the
QE condition (see Fig.~\ref{fig:BHNS_XCTS_constraints_set}). A similar
diagnostic quantity can be constructed by looking at the magnitude of
$\mathcal{L}_{\boldsymbol{\xi}} K_{ij}$, whose explicit expression can be
found at the end of
Appendix~\ref{sec:3plus1_reconstruction_of_riemann}. In this case, while
$\mathcal{L}_{\boldsymbol{\xi}} K_{ij} = 0$ is not enforced explicitly
(in fact, only its trace part is), since $K_{ij}$ is a tensor of a
spacetime with global helical symmetry $\boldsymbol{\xi}$,
$\mathcal{L}_{\boldsymbol{\xi}} K_{ij}$ should necessarily vanish for all
of its components as well\footnote{Note that
$\mathcal{L}_{\boldsymbol{\xi}} g_{\mu\nu} = 0$ for a Killing vector
$\boldsymbol{\xi}$ does not imply that $\mathcal{L}_{\boldsymbol{\xi}}
K_{\mu\nu}=0$.}. Of course, since $\boldsymbol{\xi}$ is a helical
symmetry vector of the spacetime only approximately, both of these Lie
derivatives are expected to be as small as possible globally for the set
of equations to be consistent with the QE assumption.

In Fig.~\ref{fig:BHNS_XCTS_constraints_set} we report the constraints
defined in Eqs.~\eqref{eq:XCTS_constraints_1}, ~\eqref{eq:XCTS_constraints_2} and
\eqref{eq:XCTS_constraints_3}, as well as the two extra diagnostics
quantities \eqref{eq:evol_eq_conf_factor} and \eqref{eq:dtKij_eq} as a
way to judge the degree of helical symmetry in the initial data for a
representative BHNS binary system. More specifically, the figure refers
to an irrotational BHNS system with a separation $d = 56.865\,{\rm km}$,
where the NS mass (in isolation) is $M_{_{\rm NS}} = 1.300\, M_{\odot}$
and the BH Christodolou mass is $M_{\rm Ch} = 6.500\, M_{\odot} =
5\,M_{_{\rm NS}}$. The distance between the objects was chosen so as to
show the level of constraint violations at a close proximity as measured
by $d/M_{\rm tot}=4.94$. The spectral resolution used here and throughout
the paper is $N = 13$ and the definition of this spectral resolution will
be discussed in Sec.~\ref{sec:methods_and_parameter_space}.

While the constraint equations are solved to a high degree of precision
at each spectral collocation point,
Fig.~\ref{fig:BHNS_XCTS_constraints_set} shows the violations obtained
from the spectral interpolation function for each constraint
equation. Note that the interpolants for $(\mathcal{H}, \mathcal{M}^{i},
\mathcal{C}_{\alpha\psi})=\bm{0}$ are satisfied at an average absolute
precision of $10^{-10}-10^{-9}$, with violations that increase by about
one order of magnitude in the vicinity of the BH, and reaching at most
$\sim 10^{-6}$ around the NS. The degree of ``non-helicity'' as measured
by $\mathcal{L}_{\boldsymbol{\xi}} \ln \psi$ and
$\mathcal{L}_{\boldsymbol{\xi}} K_{xx}$ shown in the two rightmost panels
is consistent with previous studies under similar QE
assumptions~\cite{Cook:2004kt, Foucart2008}. Overall, the level of constraint
violations shown in Fig.~\ref{fig:BHNS_XCTS_constraints_set} is
characteristic for all of our initial data configurations and is optimal
in the sense that decreasing it further would only be possible with a
considerable additional computational cost by going to higher spectral
resolution. Conversely, going to higher spectral resolution can also lead
to introducing spectral noise (i.e. Gibbs phenomenon) at the NS surface
which can spoil the solution; therefore, a conservative spectral
resolution is necessary without resorting to spectral filtering algorithms.
Finally, because the example considered refers
to a binary configuration at a very close separation distance, it
represents an upper bound for the level of constraint violations for the
binary configurations reported in this paper.

\begin{figure}[t!]
  \includegraphics[width=\columnwidth, keepaspectratio]{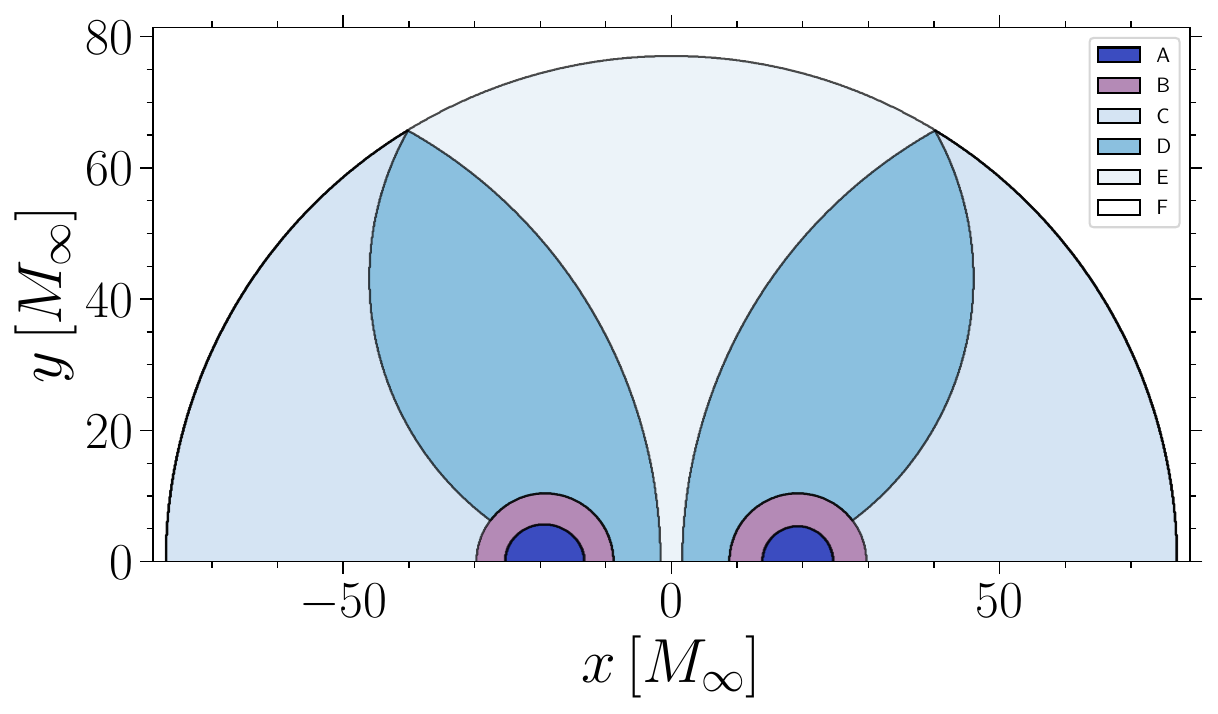}
  \caption{Representative example of the non-overlapping domain
    decomposition used in \fuka for constructing BHNS initial data. The
    legend for the various patches is described in the text and the
    figure is generated using the same BHNS binary described in
    Fig.\ref{fig:BHNS_XCTS_constraints_set}.}
  \label{fig:domains}
\end{figure}

As a final remark and mostly for completeness, it is instructive to
mention that conformally flat initial data with a spinning isolated BH is
inherently incompatible with the axisymmetric slicing of Kerr
spacetime~\cite{Garat:2000pn, Kroon2004, DeFelice2019} and effectively
introduces unphysical radiation content. Techniques have been reported in
the literature when constructing binary initial data that includes a BH
which abandon the conformally flat approximation and, instead, compute
conformally curved initial data where $\tilde{\gamma}_{ij} \neq
\delta_{ij}$ and $K \neq 0$. Conformally curved BHNS initial data has
previously been constructed using the modified-Kerr-Schild conformal
metric~\cite{Foucart2008, Foucart2010, Tacik:2016zal}, based on the
previous super-imposed Kerr-Schild data~\cite{Lovelace2008c} for BBH
systems. While Ref.~\cite{Taniguchi2006} used the Kerr-Schild
representation of the Schwarzschild metric, only irrotational sequences
were considered there. Finally, it is worth noting that although
techniques exist in the literature for constructing conformally curved
initial data, no public code exists with these capabilities with the
exception being the elliptic solver in \texttt{SpECTRE} for generating
conformally curved BBH initial data~\cite{Vu2021a, Vu2024a}.

\section{Numerical setup}
\label{sec:methods_and_parameter_space}

\subsection{The \fuka code}

The results presented here have been obtained with the open-source and
publicly available \fuka code, which is a suite of initial-data solvers
designed to compute initial conditions for isolated and binary compact
objects, with a focus on highly mass asymmetric and spinning
configurations. \fuka is built on an extended version of the \kadath
spectral-solver library which is a highly-scalable,
\texttt{MPI}-parallelized, \texttt{C++} library specifically designed for
numerical relativity problems~\cite{Grandclement09}. To discretize the
physical space, \kadath utilizes a multi-grid approach where multiple
non-overlapping grids are used which are chosen based on the problem
being solved. Figure~\ref{fig:domains} illustrates the domain
decomposition for BHNS initial data. Regions A (right) denote the excised
region of the BH or (left) the stellar interior. The boundaries between
regions A and B are adaptive and are not required to remain spherical.
Instead, they are specifically designed to adapt to the surface of a
problem which, in this case, is the apparent horizon radius dictated by
the irreducible mass of the BH or the
deformation of the NS surface. It is important to note that the latter is
sensitive not only to the mass and spin of the NS, but also the tidal
deformations that result from computing initial data with small
separation distances (\ie $d/M_{\rm tot} \lesssim 6$). Secondly, although
not shown for this particular initial data, region B can include
additional spherical shells to provide local refinement without
increasing the spectral resolution. This is extremely important for BHNS
systems with $M_{\rm Ch} \ll 5\,M_{_{\rm NS}}$ and $|\chi_{_{\rm BH}}|
\gg 0$ for which the excised region becomes increasingly smaller than the
NS radius~\cite{Markin2023, Tootle2024a}. Region F consists of the outermost
spherical shell that utilizes a compactified radial
coordinate. Finally, regions C, D, and E form a bispherical decomposition
that provides a continuous chart between the outermost spherical
boundary of each compact object and the compactified outer domain.
Currently, each spectral domain consists of the same number of
collocation points in the $r$, $\theta$, and $\phi$ directions. Here, we
define a resolution $N$ such that the spectral resolution in ($r$,
$\theta$, $\phi$) is ($N$, $N$, $N-1$).

\begin{figure*}
  \includegraphics[width=0.45\textwidth]{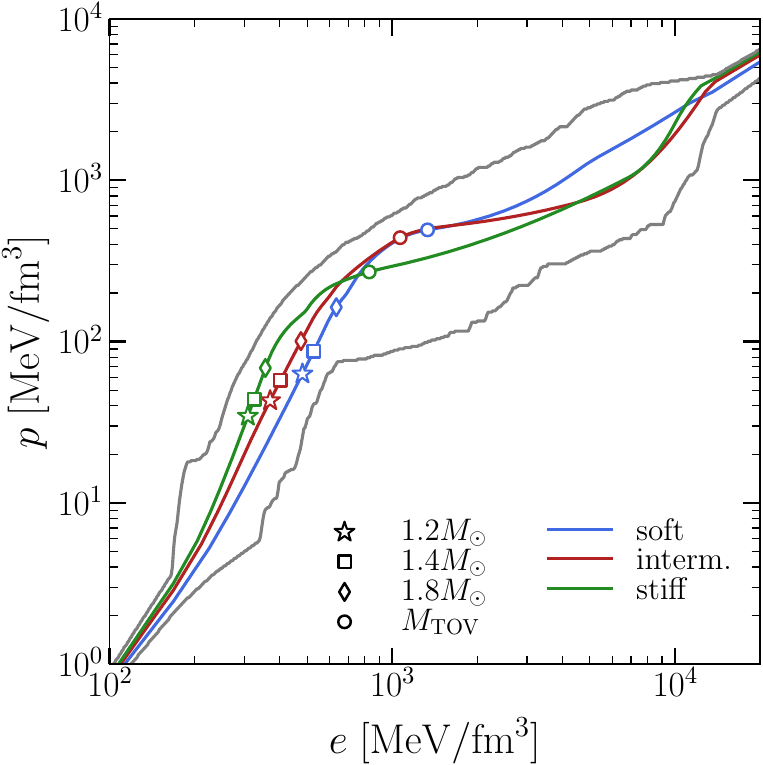}
  \hskip 0.5cm
  \includegraphics[width=0.45\textwidth, height=0.505\textwidth]{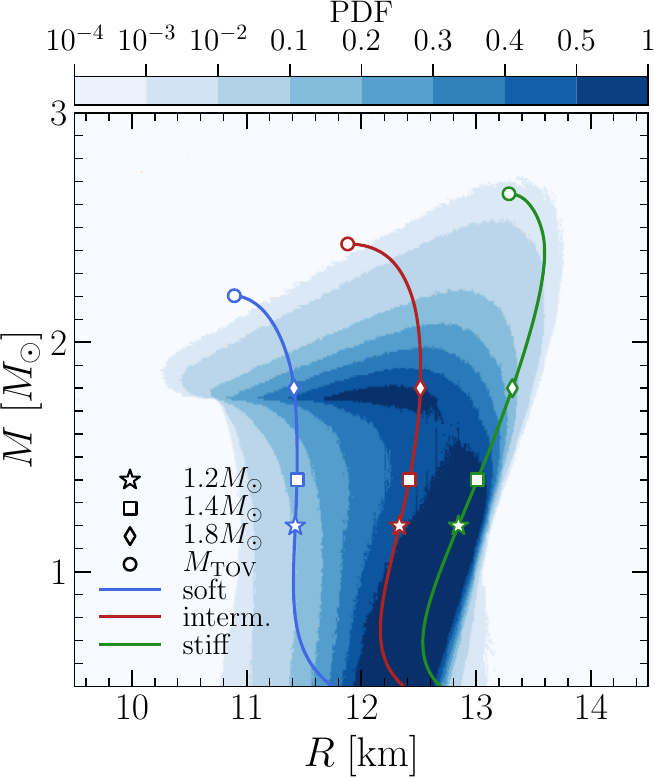}
  \caption{\textit{Left panel:} the pressure and energy density profiles for the
    soft (blue solid line), intermediate (red solid line) and stiff (green solid
    line) EOSs, respectively, utilized in this work. Shown instead with grey
    contours are the ranges of pressures and energy densities within which the EOSs
    are admissible under chiral-field theory, perturbative quantum chromodynamical
    (QCD) constraints, and astronomical observations. \textit{Right panel:} the
    same EOSs as in the left panel, but shown in terms of the $M$-$R$ curves for
    nonrotating stellar models. }
  \label{fig:eos_choice}
\end{figure*}

The system of elliptic equations is solved across all domains
consistently using a standard Newton-Raphson method where each iteration
attempts to minimize the residual error. Elliptic solvers that utilize
Jacobian methods tend to suffer from convergence issues as convergence is
heavily dependent on the quality of the initial ``guess'' used for the
first iteration. \fuka takes an iterative complexity approach to build
the initial guess from boosted isolated solutions using the XCTS system
of equations to maximize the likelihood of
convergence~\cite{Tootle2024a}. Additionally, the boosted isolated
solutions provide valuable information as to whether additional spherical
shells in region B are needed to ensure sufficient resolution is
available to resolve small features, such as the BH apparent horizon,
without resorting to globally increasing the spectral resolution. At the
same time, the surface of the NS is challenging to model with spectral
methods as it introduces a natural discontinuity that is a common source
of spectral noise known as ``Gibbs phenomenon''. Here, we minimize the
spectral noise in the numerical methods in two ways. First, \kadath
implements a non-overlapping multi-grid scheme and, therefore, the
spectral solution in a given domain describes only the local solution
within that domain with appropriate boundary conditions ensuring the
solution in one domain matches the solution in the neighbouring
domain. Secondly, we use a conservative resolution of $N = 13$, which has
shown to provide robust solutions (see
Fig.\ref{fig:BHNS_XCTS_constraints_set}) without introducing excessive
spectral oscillations at the NS surface. We have checked that the
quantities needed for the analysis as well as general diagnostics are
already robust at this resolution and no notable change is present when
the resolution is increased. The overall convergence properties of \fuka
have been demonstrated previously in ~\cite{Papenfort2021}. For the
quantities needed in our current analysis, the differences measured when
increasing the resolution to $N = 15, 17, 19$ are below $0.5\%$ for
$\kappa$ and $0.1\%$ for $E_{\rm b}$ (see below for a definition). The
differences at $N=13,15,17$ computed with respect to $N=19$ also decrease
with resolution, signalling convergence.

Furthermore, it is worth mentioning that due to the choice of spectral bases
in \kadath, only spins aligned or anti-aligned with the orbital angular
momentum are possible. It has been found, however, that the crucial
quantity in determining the outcome of the merger is the projection of
the BH spin vector orthogonal to the orbital plane, \ie $\bm{\chi}_{_{\rm
    BH}} \cdot \boldsymbol{\hat{z}$}~\cite{Foucart2010, Foucart2018}. For
that reason, we expect the results we obtained using
$\boldsymbol{\chi}_{_{\rm BH}} = \chi_{_{\rm BH}} \boldsymbol{\hat{z}}$
to be qualitatively robust and representative even for more generic
configurations. We leave it to future work to explore the potential
impact misaligned spins may have when analysing equilibrium initial data
sequences in a similar manner.

Finally, to ensure a comprehensive and model-agnostic coverage of the parameter
space of EOSs describing NS matter, we choose three representative EOSs
from an ensemble generated by uniform sampling with the speed-of-sound
parametrization (see~\cite{Altiparmak:2022, Most2018} for details). The
radius measurements of NICER~\cite{MCMiller2019b, Riley2019, Miller2021,
  Riley2021} and the upper bound on the binary tidal deformability as
measured for GW170817 by the LIGO/Virgo
collaboration~\cite{LIGOScientific:2018hze} have been imposed to reject
EOSs that violate these measurements. In the left panel of
Fig.~\ref{fig:eos_choice} we show the pressure as a function of energy
density $p(e)$ for these three EOSs. Each colour corresponds to a
different stiffness of the EOS, where blue is for the soft EOS, green is for
the stiff EOS and red is for the EOS of intermediate stiffness.
Furthermore, pressure and energy density values at the centre of the NS
$(p_{\rm c}, e_{\rm c})$ corresponding to ADM masses of the NSs
of $1.2, 1.4, 1.8\,M_{\odot}$, as well as to the
Tolman-Oppenheimer-Volkoff mass $M_{\rm TOV}$ representing a nonrotating
star of maximal admissible ADM mass are marked with symbols. In addition,
grey contours are added indicating the ranges of pressures and energy
densities where the EOSs are admissible under chiral-field theory,
perturbative quantum chromodynamical (QCD) constraints, and astronomical
observations. Clearly, the EOSs chosen roughly trace the outer contours
of the probability distribution of said ensemble, as well as the highest
likelihood region and are collectively presented in the right panel of
Fig.~\ref{fig:eos_choice} (see Ref.~\cite{Ecker:2024b} for a more
sophisticated approach for the selection of the EOSs by exploiting a
principal-component analysis of the stellar properties).

\subsection{Mass-shedding frequency and binding energy}
\label{subsec:ms_isco_measurement}

The sequences of QE binaries we consider here are characterized by
constant values of the NS baryon rest mass $M_{\rm b}$, the dimensionless
spins of the two components $\chi_{_{\rm BH/NS}}$, a fixed EOS, and
by the BH Christodolou mass
\begin{align}
  M_{\rm Ch}^2 := M_{\rm irr}^2 + \mathcal{S}^2 / 4M_{\rm irr}^2 \,;
  \nonumber
\end{align}
where $M_{\rm irr}$ is the irreducible mass of the BH and $\mathcal{S}$
the spin angular momentum measured on the apparent horizon.
In \fuka, the BH is excised from the computational domain and the
excision boundary is enforced to be an apparent horizon. As such, the
Christodolou mass provides a robust measurement of the BH gravitational
mass even in a binary configuration as the spin angular momentum and the
irreducible mass have well defined geometric
definitions~\cite{Papenfort2021}. A NS, however, does not have a
well-defined measurement of its gravitational mass in the presence of a
companion object. Instead, the ADM mass of the NS is computed and fixed
at infinite separation (\ie in isolation) and, thereafter, the baryon
rest mass will remain constant along the sequence, mimicking dynamical
evolution through QE slices.
Individual datasets along a sequence are parameterized either by the
coordinate separation or by the orbital angular velocity, which is part
of the solution.
With $M_{\rm b}$ constant and thus always representing the same star,
each sequence is also uniquely defined by the value of the NS compactness
$\mathcal{C}:=M_{_{\rm NS}}/R_{_{\rm NS}}$ as computed in
isolation\footnote{Sequences of constant quasi-local ADM mass for the NS
do not represent a useful option due to systematic
deviations~\cite{Tichy2019}. Here, $R_{_{\rm NS}}$ is the NS areal radius
and $M_{_{\rm NS}}$ its ADM mass. Therefore, for clarity and with no
risk of confusion, throughout the paper we shall always use $M_{_{\rm
    NS}}$ to denote the isolated NS ADM mass.}. For a BHNS configuration
at a given separation, we compute two important quantities necessary for
our analysis. Following Ref.~\cite{Taniguchi:2008a}, we first define the
\textit{mass-shedding} diagnostic as the ratio of the slope of the
log specific enthalpy $h$ along the equatorial and polar direction
\begin{align}
  \kappa := \frac{\partial_{r}\ln h \vert_{\rm eq}}{\partial_{r}\ln h
    \vert_{\rm pole} }\,.
\label{eq:mass_shedding_diagnostic}
\end{align}
Such a quantity is a sensitive indicator of the NS deformation resulting
from the presence of tidal forces. Clearly, $\kappa=1$ for a spherical
star (hence in isolation) while the formation of a cusp on the NS surface
can be taken as the limit in which $\kappa \to 0$, which would indicate
the onset of mass shedding. Because our solver uses spectral methods to
compute the initial data and hence cannot handle any sharp features, the
computed sequences do not extend to the mass-shedding value of $0$ and
reach $\sim 0.5$ at their smallest instead (a similar approach was
followed also in Ref.~\cite{Taniguchi:2008a}).

The second important quantity in our analysis is the \textit{binding
  energy}, also referred to as the \textit{effective potential}
(\eg~\cite{Cook94}), defined as
\begin{align}
	E_{\rm b}:= M_{_{\rm ADM}} - M_{\rm tot }\,,
\end{align}
where $M_{_{\rm ADM}}$ is the ADM mass of the hypersurface calculated
according to Eq.~\eqref{eq:adm_mass} and we define the (isolated) total
mass simply as
\begin{equation}
M_{\rm tot } := M_{_{\rm NS}} + M_{\rm Ch}\,.
\end{equation}
In analogy to the role played by the effective potential in determining
the existence and stability of circular orbits in BH spacetimes, the
minimum of the binding energy provides a working definition for the
location of the \textit{marginally stable} orbit in the spacetime
relative to the BHNS binary. In this work it was often found
necessary to extrapolate $E_{\rm b}$ to locate this minimum. This point
and its implications will be discussed in detail in
Sec.~\ref{sec:results}.

\subsection{Tidal forces analysis in the initial data}
\label{subsec:tidal_forces_ID}

Next, we briefly present a new approach to study tidal forces in
numerically generated spacetimes that is complementary to the one presented
in the previous section in terms of the mass-shedding quantity $\kappa$.
To the best of our knowledge, and leaving aside the analytical
approaches discussed in Refs.~\cite{Wiggins00, Ishii2005, Pannarale2010,
  Stockinger2024} which make use of tidal force analysis in a convenient
reference frame, this is the first time this approach is proposed and
employed for BHNS sequences.

The starting point of our analysis is represented by the equation of
geodesic deviation that characterises the influence of spacetime
curvature on the separation of neighbouring geodesic curves along a
congruence. Denoting the second covariant derivative along the direction
of the geodesic four-vector field $\boldsymbol{u}$ by
${D^2}/{D\tau^{2}}$, the spacetime Riemann tensor by $\boldsymbol{R}$ and
the separation four-vector by $\boldsymbol{X}$, the relevant equation
reads
\begin{equation}
  \frac{D^2 X^\mu}{D\tau^2} =
  R^{\mu}_{\;\;\alpha\beta\gamma}u^{\alpha}u^{\beta}X^{\gamma}\,,
  \label{eq:geodesic_deviation}
\end{equation}

We now introduce a tetrad, \ie a set of orthonormal vectors
$\{\boldsymbol{e}_{\hat{a}}\}$ with $g_{\mu\nu}
\boldsymbol{e}_{\hat{a}}{}^{\mu} \boldsymbol{e}_{\hat{b}}{}^{ \nu} =
\eta_{\hat{a} \hat{b}} = \rm{diag}(-1,1,1,1)$, with tetrad indices marked
by hatted Latin indices $\hat{a}=\{0,1,2,3\}$ and with $\eta_{\hat{a}
  \hat{b}}$ the Minkowski metric. Assuming now that the tetrad is
parallely propagated along the geodesic field, \ie
$\nabla_{\boldsymbol{u}} \boldsymbol{e}_{\hat{a}} = 0$, where
$\boldsymbol{e}_{\hat{0}} = \boldsymbol{u}$ and with $\hat{a}=\{1,2,3\}$
for some spacelike unit vectors $\boldsymbol{e}_{\hat{a}}$ at each event,
a useful simplification of the geodesic deviation equation is
achieved. Namely, the left-hand side simplifies to an ordinary second
derivative of the components of the deviation vector expressed in the
frame basis; in addition, the evolution of the timelike component
$X^{\hat{0}}$ becomes trivial. Indeed, for a decomposition of the
separation vector in the frame basis
\begin{equation}
\boldsymbol{X}=X^{\hat{0}}\boldsymbol{e}_{\hat{0}} +
X^{\hat{i}}\boldsymbol{e}_{\hat{i}}\,,
\end{equation}
one finds by writing the covariant derivative in full
\begin{equation}
  \nabla_{\boldsymbol{e}_{\hat{0}}}\boldsymbol{X} = \frac{d
    X^{\hat{0}}}{d\tau} \boldsymbol{e}_{\hat{0}} +
  X^{\hat{0}}\nabla_{\boldsymbol{e}_{\hat{0}}}\boldsymbol{e}_{\hat{0}} +
  \frac{dX^{\hat{i}}}{d\tau} \boldsymbol{e}_{\hat{i}} +
  X^{\hat{i}}\nabla_{\boldsymbol{e}_{\hat{0}}}\boldsymbol{e}_{\hat{i}}
  \,,
\end{equation}
where the second and the fourth term vanish. Applying the covariant
derivative a second time, thus yields
\begin{equation}
\frac{D^{2}\boldsymbol{X}}{D\tau^{2}} =
\frac{d^{2}X^{\hat{0}}}{d\tau^{2}} \boldsymbol{e}_{\hat{0}}+
\frac{d^{2}X^{\hat{i}}}{d\tau^{2}} \boldsymbol{e}_{\hat{i}}\,.
\end{equation}
The right-hand side of Eq.~\eqref{eq:geodesic_deviation} in the tetrad
basis becomes $R^{\hat{a}}_{\;\;\hat{0}\hat{0}\hat{b}}X^{\hat{b}}$, which
vanishes for $\hat{a}=0$ due to the symmetries of the Riemann tensor. The
general solution for the time component is then readily obtained as
$X^{\hat{0}}=\lambda_{1}\tau + \lambda_{2}$ for some constants
$\lambda_{1},\lambda_{2}$.

Defining now the ``tidal-force operator'' to be a 3-by-3 symmetric matrix
given by $C_{\hat{i}\hat{j}} = \boldsymbol{R}(\boldsymbol{e}_{\hat{i}},
\boldsymbol{u},\boldsymbol{u}, \boldsymbol{e}_{\hat{j}}) =
R_{\hat{i}\hat{0}\hat{0}\hat{j}}$, the remaining spatial components of
the geodesic-deviation equation in a parallely propagated frame amount to
[\cf Eq.~(3.11) in~\cite{Straumann2013}]
\begin{equation}
  \centering
	\frac{d^2 X_{\hat{i}}}{d\tau^2} = C_{\hat{i}\hat{j}}X^{\hat{j}}\,.
  \label{eq:geodesic_deviation_parallel_prop_frame}
\end{equation}

While this choice of frame is a particularly convenient one, it requires
the solution of both the geodesic equation for the first vector of the
tetrad, \ie $\boldsymbol{e}_{\hat{0}}$, and the solution of the equations
of parallel transport along $\boldsymbol{e}_{\hat{0}}$ for the remaining
vectors $\boldsymbol{e}_{\hat{a}}$ of the tetrad. Furthermore, because
the fluid motion is not geodesic as a result of pressure gradients, a
nonzero acceleration $\boldsymbol{a} := \nabla_{\boldsymbol{u}}
\boldsymbol{u}$ will be present and would manifest via additional terms
on the right-hand side of the geodesic deviation
equation~\eqref{eq:geodesic_deviation} (see, \eg~\cite{Straumann2013} for
more details). On the other hand, the failure to parallel transport the
remaining spacelike tetrad vectors in
Eq.~\eqref{eq:geodesic_deviation_parallel_prop_frame} would contribute
extra terms involving Christoffel symbols that account for the Coriolis
force due to the absence of Fermi-Walker transport and present for a
rotating reference frame. To cope with these shortcomings and develop an
approach that is better suited to our initial data, we here choose the
spatial vectors of our tetrad to be simple unit Cartesian coordinate
vectors and accordingly refer to the \textit{effective} ``tidal-force
operator'' of the \textit{generic} tetrad $\{\boldsymbol{e}_{\hat{a}}\}$
as the contraction of the Riemann tensor with the timelike vector field
of the frame defined above. In addition, rather than working with a
(spatial) rank-two tensor $C_{\hat{i}\hat{j}}$, which is dependent on the
choice of spatial tetrad vectors, we will instead look at its (spatial)
contraction $C_{\hat{i}\hat{j}} C^{\hat{i}\hat{j}}$, which is a scalar
field.

\begin{figure*}
  \centering
  \includegraphics[width=0.95\textwidth]{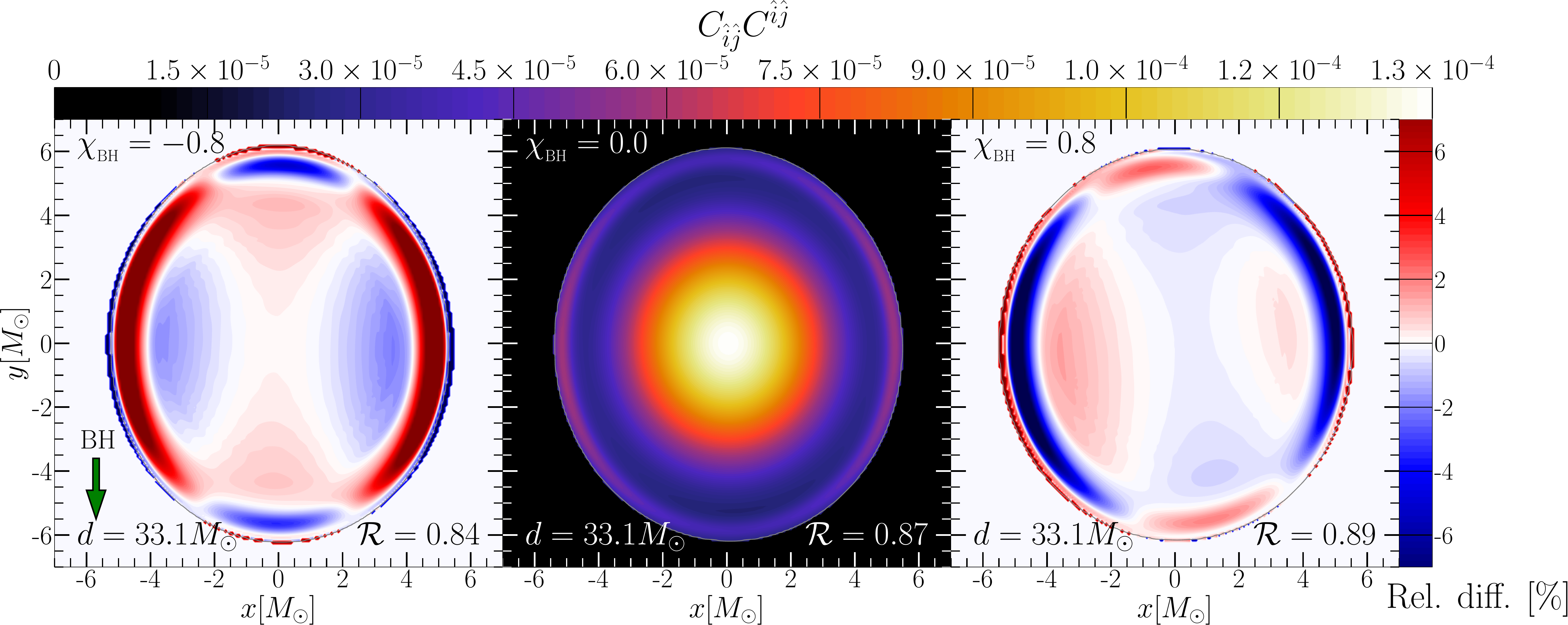}
  \vskip 0.25cm
  \includegraphics[width=0.95\textwidth]{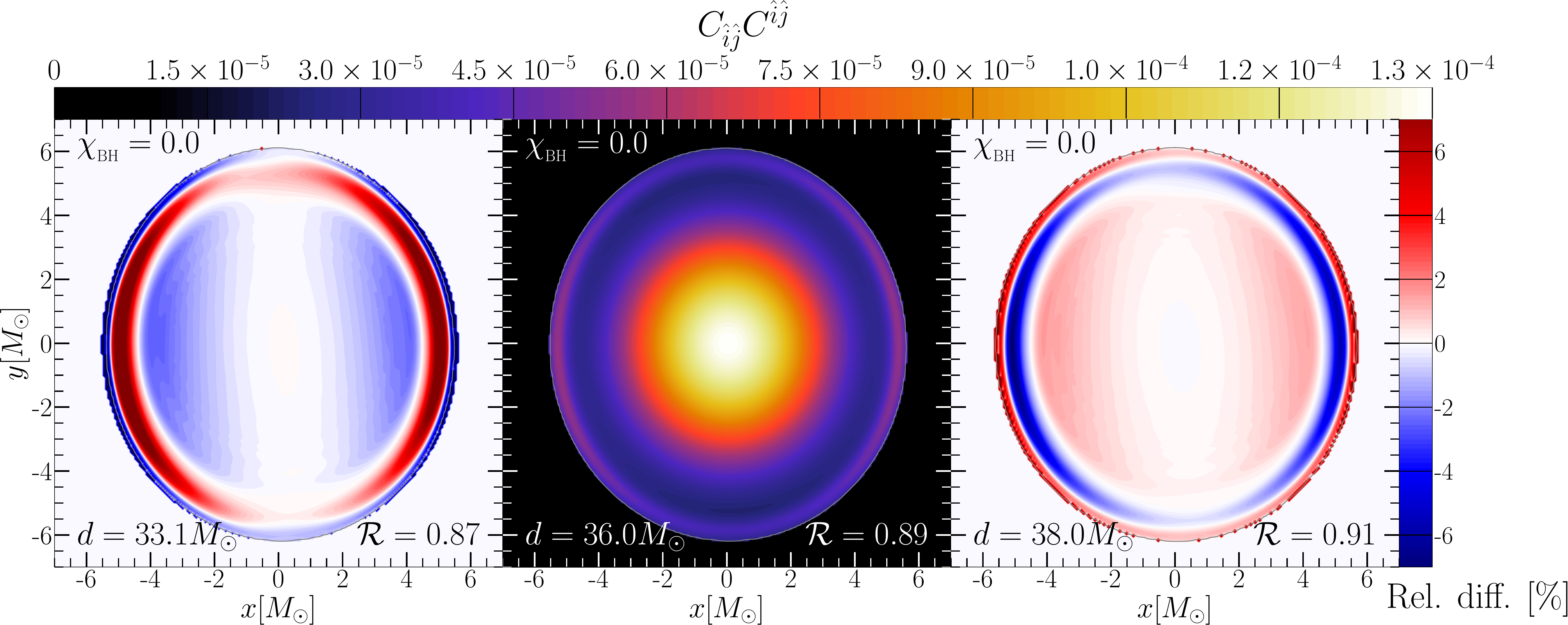}
  \caption{Representative examples of the monitoring quantity $\Upsilon
    := C_{\hat{i}\hat{j}} C^{\hat{i}\hat{j}}$, with $C^{\hat{i}\hat{j}}$
    [see Eq.~\eqref{eq:tidal_force_operator_fluid} for a definition] for
    BHNS binaries relative to a mass ratio $Q=4$, a total mass
    $M_{\rm tot}=6.640\,M_{\odot}$ and the EOS of
    intermediate stiffness; the green arrow points in the direction of
    the BH. \textit{Top panels:} tidal-force indicator $\Upsilon$ for
    different BHNS binaries having the same separation of
    $d/M_{\odot}=33.1$ but different BH spin (\ie $\chi_{_{\rm BH}}=-0.8,
    0, 0.8$, respectively). The top-left and top-right panels actually
    refer to the relative difference ${(\Upsilon - \Upsilon_{0}}) /
    ({\Upsilon + \Upsilon_{0}})$ with respect to the middle panel, with
    $\chi_{_{\rm BH}}=0$ (\ie $\Upsilon_{0}$) to help appreciate the
    differences across the sequence. Note that the largest distortions
    take place near the stellar surface and in a rather thin
    region. \textit{Bottom panels:} the same as on the top but for BHNS
    binaries having the same spin $\chi_{_{\rm BH}}=0.0$ but different
    separations (\ie $d/M_{\odot}=33.1, 36.0, 38.0$, respectively). Also
    in this case, the central panel shows $\Upsilon$ while the left and
    right panels illustrate the relative difference ${(\Upsilon -
	\Upsilon_{0}}) / ({\Upsilon + \Upsilon_{0}})$.}
  \label{fig:tidal_forces_2D_ID}
\end{figure*}

More specifically, hereafter we will consider the frame associated with
an observer comoving with the fluid and, within a 3+1 decomposition of
spacetime, we express the fluid four-velocity $u^{\mu} =
W(n^{\mu}+U^{\mu})$, with $W$ the Lorentz factor, $n^{\mu}$ the unit
time-like normal to the spacelike hypersurface, and $U^{\mu}$ the
3-velocity obtained by the projection onto the
hypersurface~\cite{Rezzolla_book:2013}. The tidal-force operator in such
a frame then reads
\begin{align}
  \label{eq:tidal_force_operator_fluid}
  C_{\hat{i}\hat{j}} &=
  \boldsymbol{R}(\boldsymbol{e}_{\hat{i}},\boldsymbol{u},\boldsymbol{u},\boldsymbol{e}_{\hat{j}})
  \\ &= W^2
  \big{[}\boldsymbol{R}(\boldsymbol{e}_{\hat{i}},\boldsymbol{n},\boldsymbol{n},\boldsymbol{e}_{\hat{j}})
    +
    \boldsymbol{R}(\boldsymbol{e}_{\hat{i}},\boldsymbol{n},\boldsymbol{e}_{\hat{k}},e_{\hat{j}})U^{\hat{k}}
    \nonumber \\ &\phantom{=} +
    \boldsymbol{R}(\boldsymbol{e}_{\hat{j}},\boldsymbol{n},\boldsymbol{e}_{\hat{k}},\boldsymbol{e}_{\hat{i}})U^{\hat{k}}
    +
    \boldsymbol{R}(\boldsymbol{e}_{\hat{i}},\boldsymbol{e}_{\hat{k}},\boldsymbol{e}_{\hat{l}},\boldsymbol{e}_{\hat{j}})U^{\hat{k}}U^{\hat{l}}
    \nonumber \big{]}\,,
\end{align}
where the third term involving a single contraction with the $n^{\mu}$
vector has been permuted using the symmetry of the fully covariant
Riemann tensor (\ie $R_{\alpha\beta\gamma\delta} = R_{\delta \gamma \beta
  \alpha}$). The necessary projections of the spacetime Riemann tensor
along the normal and spatial directions in terms of the 3+1 variables are
listed in the Appendix~\ref{sec:3plus1_reconstruction_of_riemann} along
with the reconstruction procedure.

In the top panels of Fig.~\ref{fig:tidal_forces_2D_ID} we present the
tidal-force scalar $\Upsilon := C_{\hat{i}\hat{j}} C^{\hat{i}\hat{j}}$
for BHNS binaries having different BH spins $\chi_{_{\rm BH}}$ (see top
left corner of each panel) and for a fixed separation $d$. For this
analysis we utilize the intermediate stiffness EOS, we fix the ADM mass
of the NS to $M_{_{\rm NS}}=1.328\, M_{\odot}$ and fix the mass ratio to
$Q := q^{-1} = 4$. Similarly, in the bottom panels we report the same
scalar quantity, but for binaries having $\chi_{_{\rm BH}}=0$ and
different values of the binary separation $d/M_{\odot}\in\{33.1, 36.0,
38.0\}$ and corresponding orbital angular velocities $\Omega
M_{\odot}\in\{0.01046, 0.00940, 0.00877\}$. Also reported in each panel
is the ratio of the NS semi-axes as measured in the polar direction and
in the direction of the BH $\mathcal{R}={a_{\rm pole}}/{a_{\rm eq}}$ (see
bottom right corner of each panel). Note that rather than showing
$C_{\hat{i}\hat{j}}C^{\hat{i}\hat{j}}$ for each configuration, in some of
the panels we report the normalised difference between two
configurations, \ie $(\Upsilon-\Upsilon_{0}) /( \Upsilon+\Upsilon_{0})$,
where the subscript $0$ refers to the one at zero BH spin (top panels) or
the intermediate distance (bottom panels). Using this quantity, it is
easier to appreciate that the tidal forces undergo largest changes in the
outer layers of the NS and that the maximum value of
$C_{\hat{i}\hat{j}}C^{\hat{i}\hat{j}}$, located at the centre of the NS,
increases by around $2\%$ in the counter-rotating BH case when compared
to the co-rotating one. A similar behaviour is observed in the lower
panels, where a decrease in the binary separation is clearly accompanied
by an increase in the tidal force in the outer layers of the NS,
orthogonal to the line connecting the two objects (indicated by the red
regions), with a less prominent decrease closer to the core (in
blue)\footnote{Individual components of $C_{\hat{i}\hat{j}}$ may display
different longitudinal and transverse changes, but the full contraction
$\Upsilon$ manages to provide the basic and generic behaviour.}. The
maximum of $C_{\hat{i}\hat{j}}C^{\hat{i}\hat{j}}$ grows by $0.1\%$
between the largest and the smallest separations. The axis-ratio, on the
other hand, decreases by $\sim 0.02$ between two consecutive separations.
Interestingly, when comparing across top and bottom panels in
Fig.~\ref{fig:tidal_forces_2D_ID} we observe that changing the spin from
$\chi_{_{\rm BH}}=0.8$ to $\chi_{_{\rm BH}}=-0.8$ at a fixed separation
causes a change to the NS shape of a similar magnitude as to when the
separation decreases by $\sim 7.5\,{\rm km}$. On a more formal level, we
note that the comparison made above is based on coordinate descriptions
that vary from configuration to configuration. Hence, differences in
$\Upsilon$ could be simply the result of the different gauges and not be
physical. However, given that the gauges are very similar in all the
examples considered, we expect that $\Upsilon$ depends on the used gauges
only weakly across different binaries. Hence, we expect the results
discussed to be correct, at least at a qualitative level. A more rigorous
comparison for datasets at different separations would be in principle
possible when considering points along integral curves of
$\boldsymbol{\xi}$ and where $\boldsymbol{\xi}$ is to follow
some evolutionary constraints that mimic dynamical evolution (see,
\eg~\cite{Thornburg-etal-2007a} for a related example). Such a
comparison, which, to the best of our knowledge, has not been explored so
far, would still be approximate, but it would reduce the role of local
gauges across different QE sequences.

We note that the behaviour illustrated above can be also described in
terms of the orbital angular velocity or the axis ratio. In particular,
the change in $\Omega$ is $\delta \Omega M_{\odot} = 0.00097$ between the
most extreme spins in the upper panels, while $\delta \Omega M_{\odot}
=0.00169$ when changing separations in the lower panel. At the same time,
the difference in $\mathcal{R}$ between $\chi_{_{\rm BH}}=\pm 0.8$ is
larger by $0.01$ than the difference in the lower panel, in spite of a
smaller change in $\delta \Omega$ between these configurations. In other
words, the change in separation involves a larger change in $\delta
\Omega$ and yet a smaller one in the deformation $\delta
\mathcal{R}$. This is because the tidal deformations are affected more
severely by a change in the BH spin than by the binary separation for the
inspected spins and distances. Furthermore, the greatest tidal forces and
most prominent deformation of the NS are present when the spin of the BH
is in the opposite direction to the orbital angular velocity of the
system (\ie a BH with anti-aligned spin). The analogy to non-relativistic
physics is easily made if we recall that frame dragging effectively
introduces an additional force on test particles, with the sense of
direction dictated by the BH spin. Hence, the combination of orbital
motion of the fluid with the frame dragging induced by a counter-rotating
BH is responsible for the greatest degree of deformation and the largest
tidal forces. Interestingly, in spite of a different choice for our
tetrad, the phenomenology described so far in terms of the tidal-force
scalar is consistent with the analytical predictions dating back
to~\cite{Wiggins00} and that made use of a parallely-transported tetrad
derived in~\cite{Marck83}. In the accompanying paper
II~\cite{Topolski2024c}, we will present the evolution of the scalar
$\Upsilon$ in a fully dynamical setup.

As a final remark we note that even though the inspection of tidal forces
in a suitable frame provides a considerable amount of insight into the
forces and deformations taking place within the NS matter, we will not employ
it hereafter to determine the onset of the mass-shedding or to localize
the marginally stable orbit. Rather, in order to maintain consistency with previous
literature and compare directly to published results, the remainder of
this work will employ the diagnostics presented in
Sec.~\ref{subsec:ms_isco_measurement} in terms of the dimensionless
radial deformation $\kappa$ and of the binding energy $E_{\rm b}$.

\begin{figure*}
  \hspace*{-1cm}
  \includegraphics[width=0.7\textwidth]{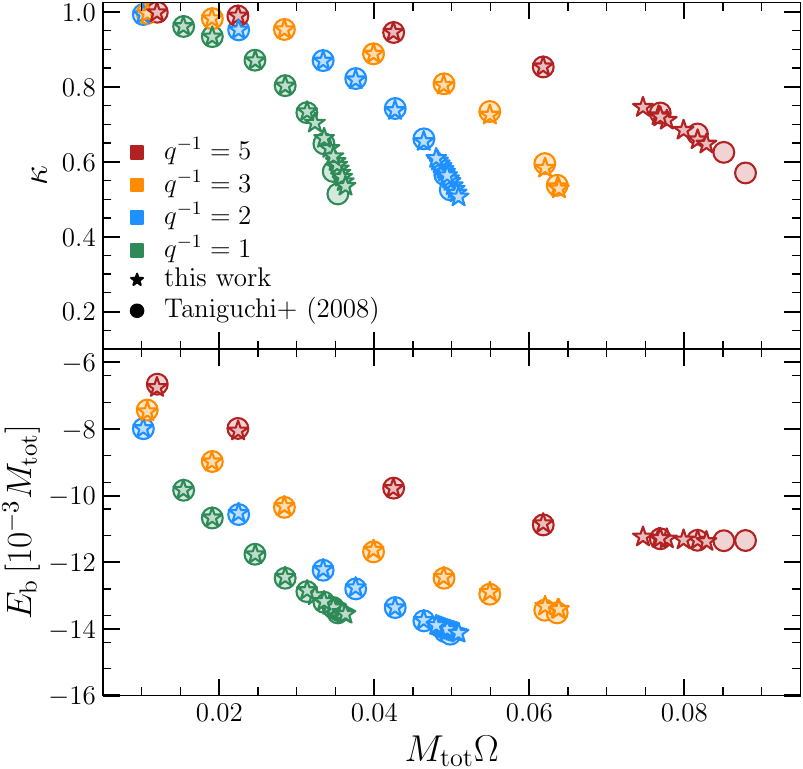}
  \caption{\textit{Top panel:} mass-shedding indicator $\kappa$ [see
      Eq.~\eqref{eq:mass_shedding_diagnostic} for a definition] along
    BHNS sequences of constant mass ratio $Q = 1,2,3,5$ and NS mass
    $M_{_{\rm NS}} =1.395\, M_{\odot}$ when using a $\Gamma=2$
    polytrope. Indicated with stars of different colours are the data
    computed here and with circles the results presented in
    Ref.~\cite{Taniguchi:2008a}. Clearly, the agreement is very good for
    most of the data points. \textit{Bottom panel:} the same as in the
    top panel but for the binding energy $E_{\rm b}$.}
  \label{fig:Gam2_Kappa_Eb_Tan08_and_T24}
\end{figure*}

\section{Results}
\label{sec:results}

In what follows, we present our results for QE sequences of BHNS binaries in
quasi-circular orbits and from which we will estimate the values of the
mass-shedding frequency $\Omega_{_{\rm MS}}$ and of the innermost stable
circular orbit $\Omega_{_{\rm ISCO}}$ as defined in
Sec.~\ref{sec:methods_and_parameter_space}.

\subsection{Polytropic EOS and comparison with literature}
\label{subsec:gam2_comparison}

Before we embark on the discussion of QE sequences using realistic EOSs,
it is useful to test the performance of the \fuka code by comparing its
results with those published in the literature. In this case, the most
exhaustive analysis is that offered by Taniguchi and
collaborators~\cite{Taniguchi:2008a}, who have considered QE sequences of
BHNS binaries when the neutron-star matter is modelled with a simple
polytropic EOS with adiabatic index $\Gamma=2$ (\ie an $n=1$ polytrope).
For the purpose of comparison, we fix the ADM masses of the NSs (and thus
the compactnesses) to those employed in~\cite{Taniguchi:2008a}. The
values of the used ADM masses in isolation and those of the baryonic
rest-masses, of the areal radii, compactnesses, and the central values of
the rest-mass densities and pressures are listed in
Tab.~\ref{tab:eos_fuka_neutron_star_diagnostics}. The mass ratio is
chosen depending on the compactness of the constant rest-mass sequence
and ranges between $Q=1$ and $Q=10$. The rest-mass values are
uniformly spaced between $1.3\, M_{\odot}$ and $1.7\, M_{\odot}$.

\begin{table}
\begin{tabular}{l|c|c|c|c|c|c}
  EOS & $M_{_{\rm NS}}$ &   $M_{b}$       & $R_{_{\rm
      NS}}$ & $\mathcal{C}$ & $10^{3} \times \rho_{c}$         & $10^{4}
  \times p_{c}$                         \\
  & $[M_{\odot}]$               &   $[M_{\odot}]$ & $[{\rm km}]$     &               & $[M_{\odot}^{-2}]$  &   $[M_{\odot}^{-2}]$  \\
  \hline \hline
  & $1.223$ & $1.300$ & $15.050$ & $0.120$ & $0.914$ & $0.835$ \\
  & $1.310$ & $1.400$ & $14.658$ & $0.132$ & $1.073$ & $1.152$ \\
 $\Gamma=2$ & $1.395$ & $1.500$ & $14.210$ & $0.145$ & $1.256$ & $1.578$ \\
  & $1.478$ & $1.600$ & $13.650$ & $0.160$ & $1.495$ & $2.236$ \\
  & $1.560$ & $1.700$ & $12.944$ & $0.178$ & $1.893$ & $3.582$ \\
  \hline
  &    $1.200$    & $1.311$ & $11.301$ & $0.157$  & $1.295$  & $1.835$ \\
  soft & $1.400$  & $1.557$ & $11.339$ & $0.182$  & $1.402$  & $2.524$ \\
  &    $1.800$    & $2.080$ & $11.345$ & $0.234$  & $1.633$  & $4.743$ \\
  \hline
  &    $1.200$    & $1.303$ & $12.184$ & $0.146$  & $1.005$  & $1.248$ \\
  intermediate    & $1.300$ & $1.423$  & $12.240$ & $0.157$  & $1.043$  & $1.442$ \\
  &    $1.400$    & $1.545$ & $12.300$ & $0.168$  & $1.083$  & $1.667$ \\
  &    $1.800$    & $2.052$ & $12.429$ & $0.214$  & $1.247$  & $2.903$ \\
  \hline
  &    $1.200$    & $1.299$ & $12.688$ & $0.140$  & $0.845$  & $0.998$ \\
  stiff &$1.400$  & $1.538$ & $12.875$ & $0.161$  & $0.885$  & $1.275$ \\
  &    $1.800$    & $2.038$ & $13.220$ & $0.201$  & $0.957$  & $1.985$ \\
   \hline
\end{tabular}
\caption{Properties of the isolated NS described with the $\Gamma=2$
  polytrope and the three realistic EOSs considered in this work.
	Listed for reproducibility are the ADM mass $M_{_{\rm NS}}$,
  the baryonic mass $M_{\rm b}$, the areal surface radius $R_{_{\rm
      NS}}$, the compactness $\mathcal{C}$, the central density
	$\rho_{c}$, and the central pressure $p_{c}$.}
\label{tab:eos_fuka_neutron_star_diagnostics}
\end{table}

In Fig.~\ref{fig:Gam2_Kappa_Eb_Tan08_and_T24} we compare the results of
our analysis for $\kappa$ (top panel) and $E_{\rm b}$ (bottom panel) with
those of Ref.~\cite{Taniguchi:2008a}. For brevity, only the QE sequences
with rest-mass $M_{\rm b}=1.5\, M_{\odot}$ and mass ratios
$Q=1,2,3,5$ are reported. Note that we find excellent agreement
along the sequences both in the mass-shedding indicator and the binding
energy, with differences that appear only at the highest frequencies,
where our solutions managed to reach somewhat larger values. These
differences, however, are much smaller than the ones that have been noted
in similar sequences, \eg for $\mathcal{C}=0.15$
in~\cite{Grandclement06}, where the whole sequence is slightly shifted.

Note that since the sequences end before encountering the mass-shedding
frequency (the iterative steps of the solution fail to converge at the
highest orbital frequencies), an extrapolation of the computed data is
necessary to extract $\Omega_{_{\rm MS}}$. To this end we resort to a
functional fitting that ensures the validity of the asymptotic limit at
infinite separation, \ie $\kappa(\Omega)=1$ in the limit of $\Omega \to
0$ instead of extrapolating from the last three data points of $\kappa$
with a quadratic polynomial as is suggested in~\cite{Taniguchi:2008a}.
More specifically, we perform a global fit of the computed data with a
functional ansatz of the type
\begin{equation}
  \label{eq:kappa_ansatz}
  \kappa_{\rm fit}(\Omega)=1 + a_1 \, \Omega^{2} \sinh (-a_2 \,
  \Omega)\,,
\end{equation}
where data points are not weighted equally in the fit, but with a growing
weight $w(\Omega)=\Omega$ to favour the points at the interesting end of
the sequence, and set the mass-shedding frequency from the condition
\begin{equation}
\kappa_{\rm fit}(\Omega_{_{\rm MS}})=0\,.
\end{equation}
In this way, we find that for the less challenging configurations this
ansatz~\eqref{eq:kappa_ansatz} provides almost identical results to the
usual extrapolation with a quadratic function on the last three points
and meaningful (\ie not excessively large) values of $\Omega_{_{\rm MS}}$
even if the sequence ends at $\kappa \approx 0.75-0.8$.

\begin{figure*}
  \center
  \includegraphics[width=0.44\textwidth]{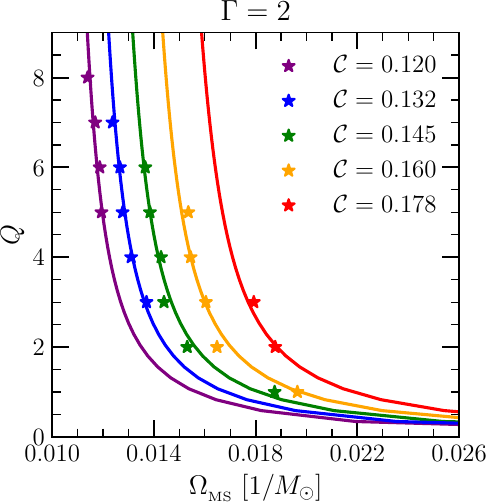}
  \hskip 0.5cm
  \includegraphics[width=0.44\textwidth]{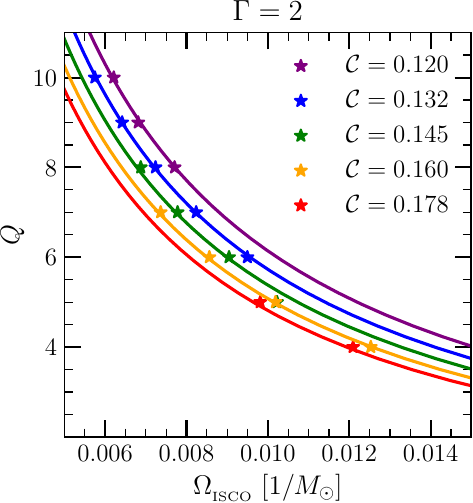}
  \caption{\textit{Left panel:} shown with coloured symbols are the
    extrapolated values of the orbital angular velocity at the onset of
    mass shedding along BHNS sequences of constant compactness
    $\mathcal{C}$. Solid lines of the same colour report the fit with the
    functional form given by
    Eq.~\eqref{eq:mtotomega_fits_ms}. \textit{Right panel:} the same as
    on the left but for the orbital angular velocity at the ISCO; also in
    this case, the solid lines show the fit given by
    Eq.~\eqref{eq:mtotomega_fits_isco}.}
  \label{fig:Gam2_Omegas}
\end{figure*}

It is often the case that the calculation of the ISCO frequency also
requires a functional fitting and extrapolation as most of the QE
sequences do not always reach the orbital frequencies where a minimum of
the binding energy can be measured. This happens, for instance, for QE
sequences having small values of $Q$, where cusp formation starts
before the ISCO is encountered, \ie $\Omega_{_{\rm MS}} < \Omega_{_{\rm
    ISCO}}$ (see the sequence with $Q=2$ in
Fig.~\ref{fig:Gam2_Kappa_Eb_Tan08_and_T24}). On the other hand, the
$Q=5$ sequence clearly ends with a measurable minimum in the binding
energy. Hence, we choose a functional ansatz for the binding energy that
respects the asymptotic limit, \ie $E_{\rm b}(\Omega)=0$ in the limit of
$\Omega \to 0$
\begin{equation}
  \label{eq:Eb_ansatz}
  E_{\rm b, fit}(\Omega)= b_1\, \Omega + b_2\, \Omega^{2} \,,
\end{equation}
and define the ISCO frequency from the condition
\begin{equation}
\left. \frac{\partial E_{\rm b, fit}}{\partial \Omega}
\right\vert_{\Omega_{_{\rm ISCO}}}= 0\,.
\end{equation}
Examples of the fitting functions~\eqref{eq:kappa_ansatz} and
\eqref{eq:Eb_ansatz} can be seen as solid gray lines in the two panels of
Fig.~\ref{fig:bhns_seq_with_spin_qhat5}, which we will discuss further
below.

\subsection{$\Omega_{_{\rm MS}}$ and $\Omega_{_{\rm ISCO}}$: dependence on $q$, and $\mathcal{C}$}
\label{subsec:analytical_formulae_MS_ISCO}

In this section we present an intermediate investigation of the
functional dependencies $\Omega_{_{\rm ISCO}} = \Omega_{_{\rm ISCO}} (Q,
\mathcal{C})$ and $\Omega_{_{\rm MS}} = \Omega_{_{\rm MS}} (Q,
\mathcal{C})$ [we recall that $Q:=M_{_{\rm BH}}/M_{_{\rm NS}}=1/q$] based
on the initial results presented in Sec.~\ref{subsec:gam2_comparison} and
illustrated in Fig.~\ref{fig:Gam2_Kappa_Eb_Tan08_and_T24}. This analysis
will be continuously elaborated upon and extended in the following
sections as we detail the results from our investigation to determine the
orbital angular velocities at the ISCO $\Omega_{_{\rm ISCO}}$ and at
mass-shedding $\Omega_{_{\rm MS}}$ for a variety of mass ratios and
compactnesses.
Our analysis of the irrotational QE sequences suggests a simple
analytical form for $\Omega_{_{\rm ISCO}} (Q,\mathcal{C})$ and
$\Omega_{_{\rm MS}} (Q, \mathcal{C})$ for which we propose the following
expressions
\begin{align}
  \label{eq:mtotomega_fits_ms}
  \mathbin{\color{blue!90!black}{M_{\rm tot}}}\, \Omega_{_{\rm MS}} (Q,
  \mathcal{C})& :=
	c_{1}\, \mathbin{\color{blue!90!black} \mathcal{C}^{3/2}
    \big{(}1+Q\big{)} \big{(}1+1/Q \big{)}^{1/2} }
  \nonumber \\
  &\phantom{=} \phantom{c_{1}\,} \times (1+c_{2}\,\mathcal{C})(1+c_{3}\,Q) \,, \\
  \label{eq:mtotomega_fits_isco}
	\mathbin{\color{blue!90!black}{M_{\rm tot}}}\, \Omega_{_{\rm
            ISCO} } (Q, \mathcal{C})& := \mathbin{\color{blue!90!black}
          6^{-3/2}} \big{(}1 + d_{1} / Q^{d_{2}} \big{)} \big{(} 1 +
        d_{1}\, \mathcal{C}^{d_{3}}/Q \big{)} \,,
\end{align}
where terms in dark blue denote necessary contributions that stem from
the test-particle limit in Kerr spacetime (for $\Omega_{_{\rm ISCO}}$) or
follow from the equality between the self-gravity force of the NS and
tidal force from the BH in Newtonian gravity (for $\Omega_{_{\rm
    MS}}$). The constants $c_{1}, c_{2}, c_{3}, d_{1}, d_{2}, d_{3}$ are
free parameters to be fitted. The terms in black in
Eqs.~\eqref{eq:mtotomega_fits_isco} and \eqref{eq:mtotomega_fits_ms} have
been introduced to refine the dependence of the mass ratio and
compactness to better match the numerical data and to properly reproduce
the relevant limits. For example, the compactness dependence in the
functional form of $\Omega_{_{\rm MS}}$ respects the limit $\Omega_{_{\rm
    MS}} = 0$ for $\mathcal{C}\to 0$, since an infinitely extended NS is
already ''disrupted'' at infinite separation. Similarly, the mass-ratio
dependence in $\Omega_{_{\rm ISCO}}$ is such that for the test particle
limit $1/Q \to 0$, $M_{\rm tot}\Omega_{_{\rm ISCO}} = 6^{-3/2}$. In
Table~\ref{tab:fitting_params_mtotomega} we list the best-fit
coefficients for the $\Gamma=2$ sequences that have been used to generate
the solid lines in Fig.~\ref{fig:Gam2_Omegas}.

\begin{table}
\begin{tabular}{l|c|c|c|c|c|c}
  EOS &  $c_{1}$  & $c_{2}$ & $c_{3}$ & $d_{1}$ & $d_{2}$ & $d_{3}$ \\
  \hline \hline
  $\Gamma=2$ & $0.334$    & $-0.218$  & $-0.003$ & $0.575$ & $0.447$ & $0.676$ \\
  \hline
  soft       & $0.221$    & $2.994$ & $-0.005$ & $0.045$ & $-1.078$  & $2.055$ \\
  \hline
  intermediate  & $0.261$ & $1.912$ & $-0.010$ & $0.126$ & $-0.277$  & $1.151$ \\
  \hline
  stiff      & $0.264$    & $1.821$ & $-0.011$ & $0.049$ & $-0.538$  & $0.720$ \\
   \hline
\end{tabular}
\caption{Best-fit parameters for the mass-shedding frequency
  $\Omega_{_{\rm MS}}$ [see Eq.~\eqref{eq:mtotomega_fits_ms}] and ISCO
  frequency $\Omega_{_{\rm ISCO}}$ [see
    Eq.~\eqref{eq:mtotomega_fits_isco}], for BHNS sequences employing
  either $\Gamma=2$ polytrope EOS or the three realistic EOSs considered
  here.}
\label{tab:fitting_params_mtotomega}
\end{table}

\begin{figure}
  \centering
  \includegraphics[width=0.96\columnwidth]{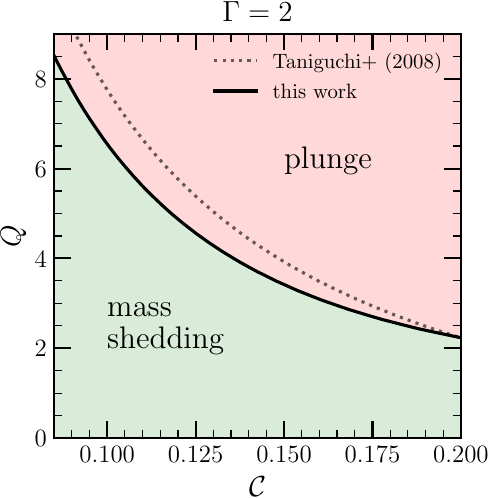}
  \caption{Shown with a solid black line is the separatrix $\Omega_{_{\rm
        MS}}(Q, \mathcal{C}) = \Omega_{_{\rm ISCO}}(Q, \mathcal{C})$ in
    the $(Q, \mathcal{C})$ variables distinguishing between BHNS binaries
    leading to a ``plunge'' (salmon-shaded area) and to a disruption
    (green-shaded area). The data refers to the sequences with a
    $\Gamma=2$ polytrope reported in Fig.~\ref{fig:Gam2_Omegas}. Shown
    with a dotted line is the corresponding separatrix suggested in
    Ref.~\cite{Taniguchi:2008a}.}
  \label{fig:Gam2_TD_vs_Plunge_Tan08_and_T24}
\end{figure}

We note that while our functional modelling of $\Omega_{_{\rm ISCO}}(Q,
\mathcal{C})$ is different from that suggested in
Ref.~\cite{Taniguchi:2008a}\footnote{In particular,
Ref.~\cite{Taniguchi:2008a} considers the ansatz $M_{\rm
  tot}\Omega_{_{\rm ISCO}} = c_{1}[ 1 - c_{2}q^{\lambda_{1}}(1 - c_{3}
  \mathcal{C}^{\lambda_{2}} ) ]$ and first fixes the coefficients
$(c_{1},c_{2},c_{3})$ by requiring compatibility with the test-particle
limit, an equal-mass BBH system and one representative BHNS binary, and
subsequently chooses the powers of $\lambda_{1,2}$ empirically to
correspond well to the data. In contrast, our fitting
function~\eqref{eq:mtotomega_fits_isco} respects by design only the
test-particle limit and constrains the free coefficients entirely from
the data.}, it yields a match to the data that is quantitatively
comparable. On the other hand, the modelling of $\Omega_{_{\rm MS}}(Q,
\mathcal{C})$ not only is functionally different, but it also leads to a
different description of our data, which is not well described by the
fitting expressions suggested in Ref.~\cite{Taniguchi:2008a}. More
specifically, we find that the additional terms in $q$ and $\mathcal{C}$
are important to capture the behaviour of our QE sequences and that
ignoring them leads to a systematic bias in the estimate of the
mass-shedding limit ($c_{2}=c_{3}=0$ in Ref.~\cite{Taniguchi:2008a}). In
Fig.~\ref{fig:Gam2_Omegas} we present the numerical data (coloured stars)
and functional forms \eqref{eq:mtotomega_fits_isco},
\eqref{eq:mtotomega_fits_ms} (coloured solid lines) along sequences of
constant compactness for BHNS binaries where the NS is modelled with a
$\Gamma=2$ EOS. Note how the fitting expressions reproduce the data very
accurately despite the very steep changes in $\Omega_{_{\rm MS}}$ and
$\Omega_{_{\rm ISCO}}$ with the mass ratio.

Finally, since we have expressions for $\Omega_{_{\rm MS}}(Q,
\mathcal{C})$ and $\Omega_{_{\rm ISCO}}(Q, \mathcal{C})$, it is
interesting to mark in the $(Q, \mathcal{C})$ space the set of points for
which $\Omega_{_{\rm MS}}(Q, \mathcal{C}) = \Omega_{_{\rm ISCO}}(Q,
\mathcal{C})$. This is shown in
Fig.~\ref{fig:Gam2_TD_vs_Plunge_Tan08_and_T24} for irrotational binaries
($\chi_{_{\rm BH}}=0$), where we report with a black solid line the
condition for which the mass-shedding and the ISCO frequencies are the
same. This separatrix essentially splits the space of parameters $(Q,
\mathcal{C})$ into two distinct regions. In the first one, indicated as
``mass shedding'' and shown with a green-shading, $\Omega_{_{\rm MS}} <
\Omega_{_{\rm ISCO}}$ and, hence, binaries here are characterised by the
tidal disruption of the NS, such that a remnant disc can be produced and
a potential electromagnetic counterpart is expected. In the second
region, indicated as ``plunge'' and shown with a salmon-shading,
$\Omega_{_{\rm MS}} > \Omega_{_{\rm ISCO}}$, implying that binaries in this
region are characterised by the NS being absorbed by the BH essentially
intact, such that no electromagnetic counterpart is to be expected.

\begin{figure*}
  \centering
  \includegraphics[width=0.42\textwidth]{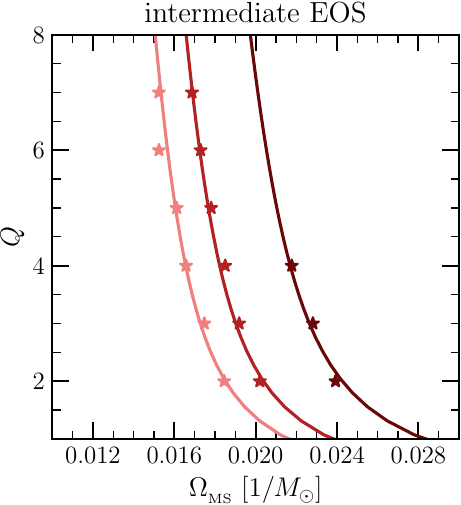}
  \hskip 0.5cm
  \includegraphics[width=0.42\textwidth]{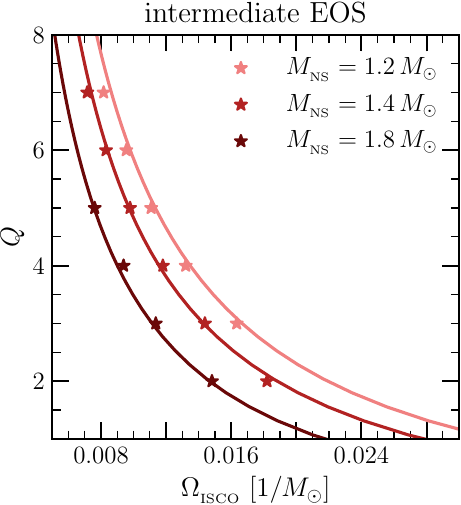}
  \caption{The same as in Fig.~\ref{fig:Gam2_Omegas} (\ie mass-shedding
    frequencies on the left and ISCO frequencies on the right), but for
    the realistic EOS with intermediate stiffness and for BHNS sequences
    of constant NS mass, \ie $M_{_{\rm NS}} = 1.2, 1.4,
    1.8\,M_{\odot}$.}
  \label{fig:Intermediate_Omegas}
\end{figure*}

\begin{figure}
  \center
  \includegraphics[width=0.99\columnwidth]{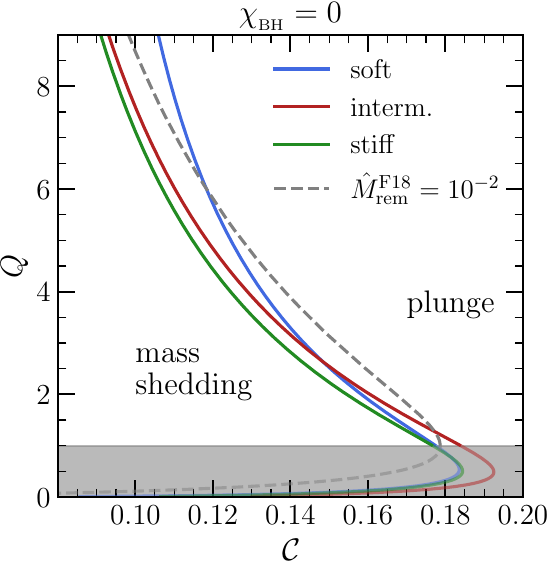}
  \caption{The same as in Fig.~\ref{fig:Gam2_TD_vs_Plunge_Tan08_and_T24},
    namely, the separatrix curves between plunge and disruption
    configurations, but for the three realistic EOSs considered here:
    soft (blue solid line), intermediate (red solid line), and stiff
    (green solid line). The BHNS binaries are irrotational (\ie
    $\chi_{_{\rm BH}}=0$) and the dashed gray line marks
    BHNS binaries with a normalized remnant mass $\hat{M}_{\rm rem}\geq
    10^{-2}$ as estimated from the fitting model of
    Ref.~\cite{Foucart2018b}. A shaded region with semi-transparent
    curves shows the very different behaviour of the separatrices for
    mass ratios $Q\leq 1$.}
  \label{fig:TD_vs_Plunge_AllEOS_and_F18}
\end{figure}

Overall, the diagram shown in
Fig.~\ref{fig:Gam2_TD_vs_Plunge_Tan08_and_T24} shows rather clearly that
mass-shedding for BHNS binaries with nonrotating BHs requires NSs that
are not very compact (\ie overall stiff EOSs) for realistic mass ratios
(\ie $Q \gtrsim 5$). Alternatively, a plunge can still take place
with smaller mass asymmetries (\ie $Q \sim 2$), but requires very
compact NSs (\ie overall soft EOSs). We will see that this behaviour
continues to apply when considering realistic EOSs (see
Fig.~\ref{fig:TD_vs_Plunge_AllEOS_and_F18}). Finally, shown with a black
dotted line in Fig.~\ref{fig:Gam2_TD_vs_Plunge_Tan08_and_T24} is the
representation of the separatrix as predicted in
Ref.~\cite{Taniguchi:2008a} and that, for the same compactness, predicts
mass shedding for larger mass asymmetries or, alternatively, for larger
compactness when holding the mass ratio constant. The difference between
the two results is small and its origin is to be found in the different
functional forms for $\Omega_{_{\rm MS}}$ and $\Omega_{_{\rm ISCO}}$, which, as
discussed above, leads to a systematic difference for binaries with small
compactness.

\subsection{Realistic EOSs: irrotational sequences}
\label{sec:bhns_no_spin}

We now turn our attention to sequences of BHNS initial data with three
realistic EOSs of varying stiffness. Instead of considering QE sequences
with constant rest-masses as was done for the polytropic EOS, we
concentrate here on astrophysically relevant values of the
gravitational mass, \ie $M_{_{\rm NS}}=1.2,1.4,1.8 \,
M_{\odot}$. Furthermore, because we consider three realistic EOSs with
different stiffness, the corresponding ranges in compactness are not the
same but overlap in the most astrophysically relevant regime
$\mathcal{C}=0.15-0.2$. Finally, we restrict the range of mass ratios
to $Q=2-7$, as realistic EOSs display greater compactnesses than the
$\Gamma=2$ polytrope (see Fig.~\ref{fig:eos_choice} and
Tab.~\ref{tab:eos_fuka_neutron_star_diagnostics}).

For brevity, we do not report here the details of the behaviour of
$\kappa$ and $E_{\rm b}$ for sequences of constant mass ratio and
NS gravitational mass, which are actually very similar to
Fig.~\ref{fig:Gam2_Omegas}. Rather, we concentrate directly on the actual
dependencies of $\Omega_{_{\rm MS}}$ and $\Omega_{_{\rm ISCO}}$ on $Q$ and
$\mathcal{C}$ for the realistic EOSs. These are reported in
Fig.~\ref{fig:Intermediate_Omegas}, along with the result of the fit for the
intermediate-stiffness EOS (analogous figures for the soft and stiff EOS
are reported in Appendix~\ref{sec:Omega_MS_ISCO_Soft_Stiff}), where the
best-fit parameters for $\Omega_{_{\rm MS}}$ and $\Omega_{_{\rm ISCO}}$ for
each realistic EOS are listed in Tab.~\ref{tab:fitting_params_mtotomega}.

When comparing Fig.~\ref{fig:Intermediate_Omegas} with the corresponding
figure for polytropes (Fig.~\ref{fig:Gam2_Omegas}) it is easy to notice
that, in particular for $\Omega_{_{\rm MS}}$, a significant difference
appears in the sequence for $M_{_{\rm NS}}=1.8 \, M_{\odot}$ when
compared to the other sequences at masses $M_{_{\rm NS}}=1.2, 1.4 \,
M_{\odot}$. The reason behind this behaviour has to be found in the fact
that stellar models with $M_{_{\rm NS}}=1.8 \, M_{\odot}$ have a
significantly larger compactness than those with $M_{_{\rm NS}}=1.4 \,
M_{\odot}$ and this increases nonlinearly the values of $\Omega_{_{\rm
    MS}}$. Obviously, this behaviour is further amplified when comparing
with the sequences of the $\Gamma=2$ polytropes and indeed, the
compactness for the $1.8 \, M_{\odot}$ NSs with the
intermediate-stiffness EOS is $\mathcal{C}=0.214$ whereas the polytropic
sequences reach $\mathcal{C}=0.1780$ at most\footnote{More precisely, the
polytropic sequence reaches $\mathcal{C}=0.178$ for a $1.56\, M_{\odot}$
NS, which is already close to the TOV mass of $1.62\, M_{\odot}$ for that
equation of state.}. For this reason we have introduced the parameter $c_{2}$
specifically to capture the steeper dependence on the compactness.

In analogy with the analysis illustrated in
Fig.~\ref{fig:Gam2_TD_vs_Plunge_Tan08_and_T24}, we report in
Fig.~\ref{fig:TD_vs_Plunge_AllEOS_and_F18} the separatrix curves for
$\Omega_{_{\rm MS}} = \Omega_{_{\rm ISCO}}$ relative to the three
different realistic EOSs considered here and indicated with lines of
different colours: green for the stiff EOS, red for the intermediate EOS
and blue for the soft EOS. Unsurprisingly, the separatrices move
systematically to larger frequencies for softer EOSs
(see \eg Fig.~\ref{fig:Soft_Stiff_Omegas}) and hence more compact NSs.
The variance, however, is rather small especially when
compared with what was observed for the simpler polytropes. Note also that
the separatrices are very similar in the range of compactnesses $0.14
\lesssim \mathcal{C} \lesssim 0.18$. This ``quasi-universal''
behaviour will be exploited in Sec.~\ref{sec:bhns_spin} to deduce some
EOS-independent considerations on the possibility of producing a massive
merger remnant.

\begin{figure*}
  \centering
  \includegraphics[width=0.65\textwidth]{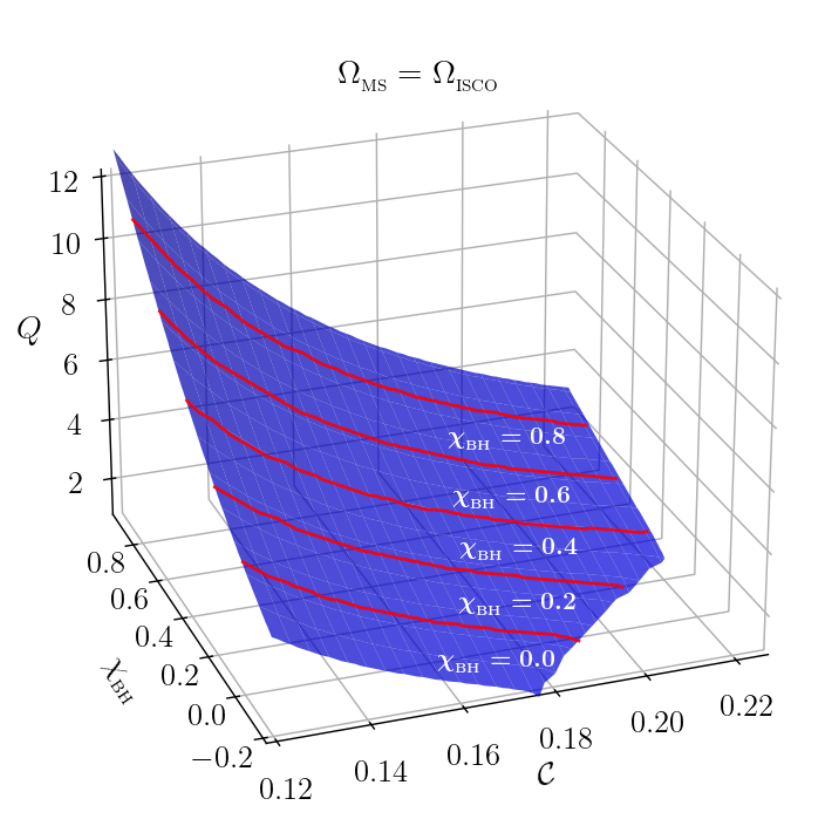}
  \caption{Shown in blue is the separatrix surface between plunge and
    disruption configurations as obtained implicitly by the condition
    $\Omega_{_{\rm MS}}=\Omega_{_{\rm ISCO}}$. Such surface represents
    the extension of the separatrix curves shown in
    Fig.~\ref{fig:TD_vs_Plunge_AllEOS_and_F18} and hence splits the space
    of parameters into regions leading to tidal disruption or
    plunge of the NS. The data refers to BHNS binaries with the
    intermediate-stiffness EOS and contours of constant BH spin are shown
    with red solid
    lines. Expression~\eqref{eq:qcrit_fit_to_omega_condition} provides a
    good fit of the functional dependence to the data on the blue
    surface.}
  \label{fig:3D_Omega_condition_separatrix}
\end{figure*}

Also shown in Fig.~\ref{fig:Gam2_TD_vs_Plunge_Tan08_and_T24} with a black
dashed line is the separatrix as computed for $\chi_{_{\rm BH}}=0$ using
the estimate for the remnant baryon mass of Foucart et
al.~\cite{Foucart2018b}. More specifically, the separatrix in this case
is given by the set of points in the $(Q, \mathcal{C})$ plane for which
the reduced remnant mass $\hat{M}_{\rm rem}=10^{-2}$ [see Eq.~(4) of
  Ref.~\cite{Foucart2018b}]\footnote{A residual mass of $\hat{M}_{\rm
  rem}=10^{-2}$ is here selected because the numerical solution of a BHNS
binary with a DD2 EOS placed on the intermediate-stiffness separatrix
would lead to a remnant disc with a mass of about $10^{-2}\,M_{\odot}$
(see paper II~\cite{Topolski2024c}).}. Considering that the remnant-mass
measurement of Ref.~\cite{Foucart2018b} combines a heterogeneous set of
numerical-relativity simulations with different EOSs, mass ratios, and BH
spins, the fact that its behaviour matches so well to the predictions
obtained from QE sequences -- which cannot account for nonlinear
dynamical effects emerging during the inspiral and disruption -- is
overall very reassuring. As we will discuss below, the predictions for
the separatrices shown in Fig.~\ref{fig:TD_vs_Plunge_AllEOS_and_F18} will
be further refined when considering the contributions to $\Omega_{_{\rm
  ISCO}}$ and $\Omega_{_{\rm MS}}$ coming from the BH spin (see
Fig.~\ref{fig:3D_Omega_condition_separatrix}).

Lastly, we note that all the separatrices in
Fig.~\ref{fig:TD_vs_Plunge_AllEOS_and_F18} show that their slopes change
sign for very small mass ratios; this behaviour, which was absent in the
case of the $\Gamma=2$ polytrope, happens around $Q \approx 0.5$ for
all the three EOSs considered in our study, and for $Q\approx 1$ in
the remnant-mass model of Ref.~\cite{Foucart2018b} (see grey-shaded
area). Overall, this scenario, which is not expected to take place under
realistic astrophysical conditions, highlights that for each EOS there
exists a maximum compactness above which a NS cannot be disrupted outside
the ISCO, no matter how small the BH mass is.

\begin{figure*}
  \centering
  \includegraphics[scale=0.7]{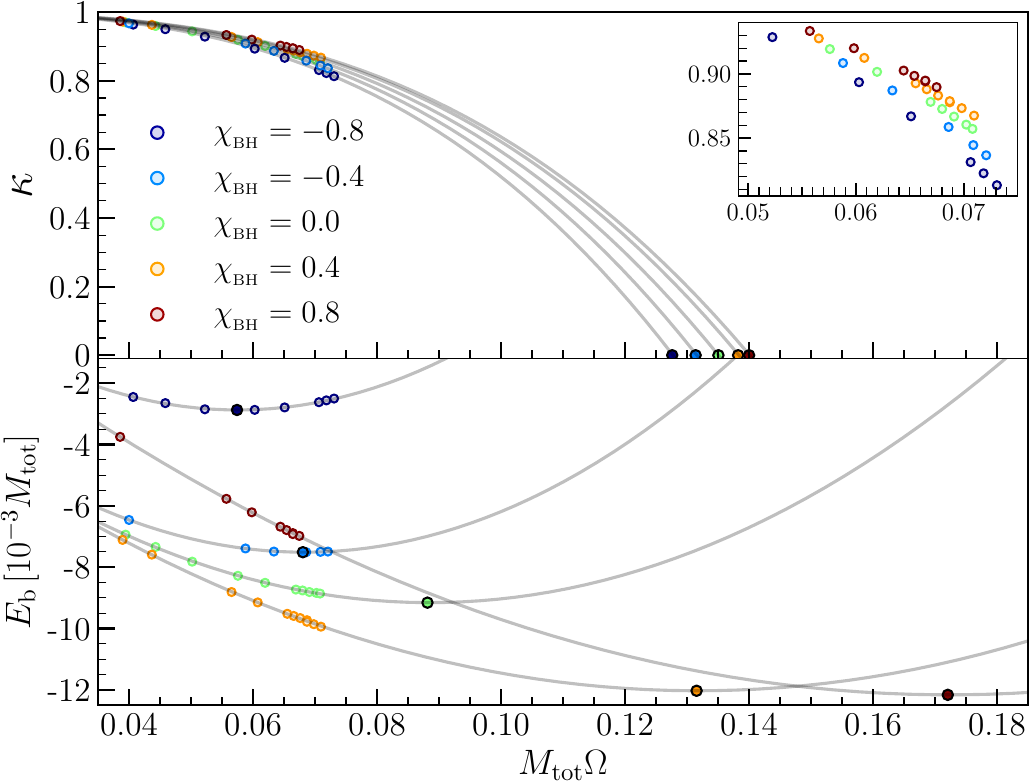}
  \hskip 0.5cm
  \caption{The same as in Fig.~\ref{fig:Gam2_Kappa_Eb_Tan08_and_T24}, but
    for QE sequences with BHs having spins $\chi_{_{\rm BH}}\in
    \{-0.8,-0.4,0,0.4,0.8 \}$. The binaries have total masses $M_{\rm
      tot}=7.8\,M_{\odot}$, a mass ratio $Q=5$ and
    employ the EOS with intermediate stiffness. Gray solid curves
    indicate the fitting functions and corresponding extrapolations
    mark the mass-shedding $\Omega_{_{\rm MS}}$ and ISCO
    frequencies $\Omega_{_{\rm ISCO}}$; the inset in the top panel
    highlights the behaviour of the numerical data.}
  \label{fig:bhns_seq_with_spin_qhat5}
\end{figure*}

\begin{table}
\begin{tabular}{c|c|c|c}
                &  $e_1$   & $e_2$    & $e_3$   \\
  \hline \hline
  $g_{_{\rm MS}}$   & $0.008$ & $-0.015$  & $0.010$  \\
  \hline
  $g_{_{\rm ISCO}}$ & $0.819$ & $0.455 $  & $--$ \\
  \hline
\end{tabular}
\caption{Fitting parameters for functions modelling the spin dependence
  defined in Eq.~\eqref{eq:mtotomega_fits_spin_terms_ms} and
  Eq.~\eqref{eq:mtotomega_fits_spin_terms_isco}, using the data presented
  in Fig.~\ref{fig:bhns_seq_with_spin_qhat5} and
  Fig.~\ref{fig:bhns_seq_with_spin_qhat4_6}.}
\label{tab:fitting_params_spin_dependence}
\end{table}

\subsection{Realistic EOSs: binaries with a spinning BH}
\label{sec:bhns_spin}

The results presented in this section follow the same methodology of the
previous sections, but specifically concentrate on QE sequences of BHNS
binaries where the BH is rotating with a different spin and is either
aligned or anti-aligned with the orbital angular momentum. Obviously,
this choice now increases the possible space of parameters, which is now
given by $(Q,\mathcal{C},\chi_{_{\rm BH}})$, making the problem harder to
tackle in general terms. For this reason, and to keep the problem
tractable, we restrict the EOS to be the one of intermediate stiffness
and the ADM mass of the NS to be $M_{_{\rm NS }}=1.300\, M_{\odot}$, so
that the corresponding baryon rest-mass and compactness is $M_{\rm
  b}=1.423\, M_{\odot}$ and $\mathcal{C}=0.168$, respectively (see
Fig.~\ref{fig:TD_vs_Plunge_AllEOS_and_F18}). In practice, this
restriction actually focuses on what is potentially the most interesting
region for realistic mass ratios $Q\approx 5$, as it leads to binaries
whose merger outcome is most affected by the presence of a spinning
BH. As a result, having fixed the EOS and the compactness, the problem
becomes much more affordable in terms of computational costs and hence we
explore $15$ QE sequences of mass ratio $Q = \{4, 5, 6\}$ and black-hole
spin $\chi_{_{\rm BH}} = \{- 0.8, -0.4, 0.0, 0.4, 0.8\}$.

Figure~\ref{fig:bhns_seq_with_spin_qhat5} provides a synthetic
representation of QE sequences with fixed values of the BH spin
and $Q=5$ in terms of the mass-shedding diagnostic $\kappa$
(top panel) and the normalised binding energy (bottom panel). Also
reported with gray solid lines are the functional fitting
functions~\eqref{eq:kappa_ansatz} and \eqref{eq:Eb_ansatz} so that it is
easy to appreciate the corresponding values of $\Omega_{_{\rm MS}}$ (\ie
$\kappa \to 0$ in the top panel) and of $\Omega_{_{\rm ISCO}}$ (\ie $\partial
E_{\rm b} / \partial {\Omega} = 0$ in the bottom panel). For compactness,
equivalent figures for mass ratios $Q=4, 6$ are presented in
Fig.~\ref{fig:bhns_seq_with_spin_qhat4_6} in
Appendix~\ref{sec:Omega_MS_ISCO_Soft_Stiff}, while the numerical values
relative to the $Q=5$ sequence are reported in
Tab.~\ref{tab:QEsequence_q5_interEOS} for reproducibility.

The spin dependence present in Fig.~$\ref{fig:bhns_seq_with_spin_qhat5}$
is clearly visible both in the mass-shedding indicator, as well as in the
binding energy. From the values of $\Omega_{_{\rm MS}}$ and $\Omega_{_{\rm
  ISCO}}$ it is apparent that both diagnostics capture the dependence on
the BH spin in a manner that is consistent with the predictions of tidal
forces in Kerr spacetime~\cite{Marck83,Wiggins00} (see also our
discussion of the tidal forces in Sec.~\ref{subsec:tidal_forces_ID}) and
the analytical dependence of the ISCO location for massive test particles
in the Kerr solution. More specifically, we find that anti-aligned BH
spins decrease systematically the values of $\kappa$ and hence yield
smaller values of $\Omega_{_{\rm MS}}$. Similarly, the BH spin affects the
slope of the $E_{\rm b}$-$\Omega$ curves to such extent that for
anti-aligned spins we can measure $\Omega_{_{\rm ISCO}}$ already among the
initial data configurations without having to resort to an
extrapolation. Interestingly, a similar behaviour has been reported also
for BH binaries with anti-aligned spins~\cite{Pfeiffer:thesis}. On the
other hand, BHs with aligned spins tend to increase the (absolute value
of the) slope of the $E_{\rm b}$-$\Omega$ curves and thus move the minima
towards higher orbital angular velocities. Again, this is exactly the
expectation from the dependence of ISCO for test-particles, for which
positive (prograde) spins tend to push the radius of the ISCO inwards,
\ie to higher orbital frequencies.

At this point, using the information from the 15 QE sequences computed,
we can express analytically and for the first time the impact the BH
spin has on the mass-shedding and ISCO frequencies. Stated differently, we
can use the bulk of data computed to derive extensions of expressions
\eqref{eq:mtotomega_fits_ms} and \eqref{eq:mtotomega_fits_isco} to
include also a dependence on $\chi_{_{\rm BH}}$.

\begin{figure*}
  \centering
  \includegraphics[width=0.99\textwidth]{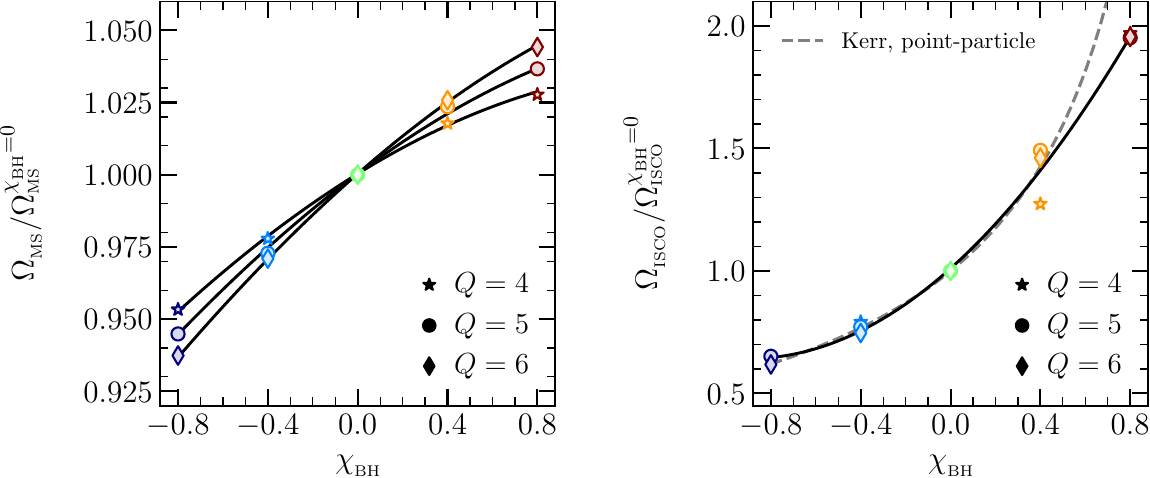}
  \caption{Orbital angular velocity at mass-shedding $\Omega_{_{\rm MS}}$
    (left panel) and at the ISCO $\Omega_{_{\rm ISCO}}$ (right panel)
    when rescaled by the corresponding values for irrotational BHNS
    binaries $\Omega_{_{\rm MS,\, ISCO}}^{\chi_{_{\rm BH}}=0}$ so as to
    remove the mass-ratio and compactness dependence. The data is
    presented with symbols of different type along sequences of mass
    ratio $Q\in\{4,5,6\}$. Shown with black solid lines are the
    results of the fits using expressions
    \eqref{eq:mtotomega_fits_spin_terms_ms} and
    \eqref{eq:mtotomega_fits_spin_terms_isco} for the various sequences
    of varying mass ratio. Shown with a gray dashed line on the right
    panel is the corresponding behaviour of $\Omega_{_{\rm ISCO}}/
    \Omega_{_{\rm ISCO}}^{\chi_{_{\rm BH}}=0}$ for a test particle around
    a Kerr BH, highlighting that deviations appear mostly for aligned and
    rapidly rotating BHs.}
  \label{fig:ISCO_MS_spin_dependence}
\end{figure*}

An effective and accurate manner to include this additional dependence is
to assume that it can be expressed as a correction to the zero-spin
expressions and hence we assume that the general expressions of
$\Omega_{_{\rm MS,\,ISCO}} = \Omega_{_{\rm MS,\,ISCO}}
(Q,\mathcal{C},\chi_{_{\rm BH}})$ have the form
\begin{align}
  \label{eq:Omega_MS_full_dependence}
	\Omega_{_{\rm MS}}(Q,\mathcal{C},\chi_{_{\rm BH}}) & := \Omega_{_{\rm
	MS}}(Q,\mathcal{C}) \, g_{_{\rm MS}}(Q, \chi_{_{\rm BH}})
  \,,\\ \Omega_{_{\rm ISCO}}(Q, \mathcal{C},\chi_{_{\rm BH}}) & :=
	\Omega_{_{\rm ISCO}}(Q, \mathcal{C}) \, g_{_{\rm ISCO}}(\chi_{_{\rm
      BH}})\,,
  \label{eq:Omega_ISCO_full_dependence}
\end{align}
where the $\Omega_{_{\rm MS,\,ISCO}}(Q, \mathcal{C})$ are the zero-spin
results given by expressions \eqref{eq:mtotomega_fits_ms} and
\eqref{eq:mtotomega_fits_isco} and discussed in
Sec.~\ref{sec:bhns_no_spin}. On the other hand, $g_{_{\rm MS}}(Q,
\chi_{_{\rm BH}})$ and $g_{_{\rm ISCO}}(\chi_{_{\rm BH}})$ are to be seen
as the additional spin-induced corrections to the relevant
frequencies. Note that in the case of the mass-shedding frequency, the
data reveals that the spin-induced corrections also depend on the mass
ratio and hence we need to express this via a function $g_{_{\rm
    MS}}=g_{_{\rm MS}}(Q, \chi_{_{\rm BH}})$; such a dependence is absent
for the ISCO frequency, so that $g_{_{\rm ISCO}}=g_{_{\rm
    ISCO}}(\chi_{_{\rm BH}})$ (see also the right panel of
Fig.~\ref{fig:ISCO_MS_spin_dependence}). At the same time, we adopt the
same spin dependence for $g_{_{\rm MS}}$ and $g_{_{\rm ISCO}}$, whose
full expressions are then given by
\begin{eqnarray}
  \label{eq:mtotomega_fits_spin_terms_ms}
	&& \hskip -0.5 cm g_{_{\rm MS}}(Q, \chi_{_{\rm BH}}) := \left ( 1 + e_1 \chi_{_{\rm
      BH}} + e_2 \chi_{_{\rm BH}}^{2} \right ) \left ( 1 + e_3
  \chi_{_{\rm BH}}Q \right)\,, \nonumber \\ \\
	&& \hskip -0.5 cm g_{_{\rm ISCO}}(\chi_{_{\rm BH}}) := \left ( 1 + e_1 \chi_{_{\rm BH}} +
  e_2 \chi_{_{\rm BH}}^{2} \right )\,,
  \label{eq:mtotomega_fits_spin_terms_isco}
\end{eqnarray}
where $(e_1, e_2, e_3)$ are free parameters to be computed by the fit and
where the functional dependence guarantees the correct zero-spin limits,
\ie$g_{_{\rm MS}}(Q, 0)=1$ and $g_{_{\rm ISCO}}(0) = 1$.

Figure~\ref{fig:ISCO_MS_spin_dependence} reports the dependence on the BH
spin of the mass-shedding frequency (left panel) and of the ISCO
frequency (right panel) when rescaled by their corresponding zero-spin
values, \ie $\Omega_{_{\rm MS, \, ISCO}}/\Omega_{_{\rm MS, \,
    ISCO}}^{\chi_{_{\rm BH}}=0}$, so as to best appreciate the
differences and provide justification for the chosen modelling
functions~\eqref{eq:mtotomega_fits_spin_terms_ms} and
~\eqref{eq:mtotomega_fits_spin_terms_isco}. More specifically, reported
with different symbols are the data points along the various constant
mass-ratio sequences while the dashed black lines are computed from the
fitting functions given by Eqs.~\eqref{eq:mtotomega_fits_spin_terms_ms}
(left panel) and \eqref{eq:mtotomega_fits_spin_terms_isco} (right panel),
respectively. The numerical values of the best-fit coefficients ($e_i$)
are reported in Tab.~\ref{tab:fitting_params_spin_dependence}. Note that
the change in the $\Omega_{_{\rm MS}}$ due to the spin is rather small
and at most around $\approx 6\%$, with a slight residual dependence on
the mass ratio. On the contrary, the BH spin significantly impacts
$\Omega_{_{\rm ISCO}}$, with $\Omega_{_{\rm ISCO}}^{\chi_{_{\rm
      BH}}=-0.8} \approx 0.6~ \Omega_{_{\rm ISCO}}^{\chi_{_{\rm BH}}=0}$
and $\Omega_{_{\rm ISCO}}^{\chi_{_{\rm BH}}=0.8}\approx 2.1~
\Omega_{_{\rm ISCO}}^{\chi_{_{\rm BH}}=0}$. Interestingly, it is possible
to compare the behaviour of the ISCO frequency to the case of a
test-particle in a Kerr-BH spacetime. This is shown as a gray dashed line
in Fig.~\ref{fig:ISCO_MS_spin_dependence} and it is quite remarkable that
the test-particle behaviour is very similar to the one we have measured
in our BHNS binaries in the case of anti-aligned spins\footnote{Note that
the normalization factor in the ISCO frequencies, \ie $\Omega_{_{\rm
    ISCO}}^{\chi_{_{\rm BH}}=0}$ is different whether one is considering
BHNS binaries or a test particle. However, once the scaling is done, the
functional behaviour of $\Omega_{_{\rm ISCO}}/ \Omega_{_{\rm
    ISCO}}^{\chi_{_{\rm BH}}=0}$ is the same for anti-aligned spins, as
shown in Fig.~\ref{fig:ISCO_MS_spin_dependence}.}. On the other hand,
significant differences appear in the case of aligned spins, where the
results of the QE sequences show a smaller increase with spin and where
the test-particle jump is $\Omega_{_{\rm ISCO}}^{\chi_{_{\rm
      BH}}=0.8}\approx 2.6~ \Omega_{_{\rm ISCO}}^{\chi_{_{\rm
      BH}}=0}$. The origin of this behaviour is to be found in the fact
that the location of the ISCO moves out to larger distances in the case
of anti-aligned spins, hence in regions of the spacetime where the
curvature is smaller and the point-particle approximation is more
accurate. On the other hand, the location of the ISCO moves in for
aligned spins, where strong-curvature and finite-size effects of the NS
are responsible for the differences seen in
Fig.~\ref{fig:ISCO_MS_spin_dependence}. Overall, these considerations
clearly highlight that the role of the BH spin is essential in
determining the location of the ISCO and hence the merger outcome. On the
other hand, the BH spin plays only a minor role in determining the
mass-shedding location.

\subsection{$\Omega_{_{\rm MS}}$ and $\Omega_{_{\rm ISCO}}$: dependence on $q$,
  $\mathcal{C}$, and $\chi_{_{\rm BH}}$}
\label{subsec:analytical_formulae_MS_ISCO+}

Having defined via expressions~\eqref{eq:Omega_MS_full_dependence} --
\eqref{eq:mtotomega_fits_spin_terms_isco} the functional dependencies of
the mass-shedding and ISCO frequencies, we can now assess the separatrix
of the ``mass-shedding'' and ``plunge'' configurations in the
three-dimensional space of parameters $(Q, \mathcal{C},\chi_{_{\rm
    BH}})$. As done in ..~\ref{fig:Gam2_TD_vs_Plunge_Tan08_and_T24}
and \ref{fig:TD_vs_Plunge_AllEOS_and_F18}, we determine this separatrix
by requiring that $\Omega_{_{\rm MS}} = \Omega_{_{\rm ISCO}}$. The
resulting surface is shown in
Fig.~\ref{fig:3D_Omega_condition_separatrix} and deserves a number of
remarks. First, we recall that while this separatrix is obtained for the
intermediate-stiffness EOS, we expect the functional behaviour to be the
same also for different EOSs. Second, it demonstrates that for a fixed
stellar compactness aligned BH spins are able to disrupt the NS at
systematically larger mass asymmetries so that, for instance, a
reference value of $\mathcal{C}=0.15$ will lead to disruption at $\chi_{_{\rm
    BH}}=0.8$ for $Q=7$, while this can be achieved for at
most $Q\approx 2.6$ for $\chi_{_{\rm BH}}=0.0$. Similarly, for a fixed
mass asymmetry, large aligned BH spins can result in a NS disruption with
systematically more compact stars so that, for instance, a reference
value of $Q=5$ will lead to disruption at $\chi_{_{\rm BH}}=0.8$ for
$\mathcal{C}=0.1718$, while this can only be achieved at compactnesses
$\mathcal{C}\approx 0.12$ for $\chi_{_{\rm BH}}=0.0$.

Because the critical separatrix given by the function $Q_{\rm
  crit}(\mathcal{C},\chi_{_{\rm BH}})$ is defined only implicitly through
the condition $\Omega_{_{\rm MS}} = \Omega_{_{\rm ISCO}}$, an explicit
analytic expression in terms of elementary functions, or possibly even a
single-branch solution in terms of special functions, is unlikely to
exist (non-integer powers of $q$ and $\mathcal{C}$ are employed in the
fitting functions~\eqref{eq:mtotomega_fits_spin_terms_ms}
and~\eqref{eq:mtotomega_fits_spin_terms_isco}). However,
we can parametrize the surface in terms of a simple fitting function
$Q_{\rm fit}$ given by the multiplication of two cubic polynomials in
$\chi_{_{\rm BH}}$ and $\mathcal{C}$
\begin{align}
  \label{eq:qcrit_fit_to_omega_condition}
	Q_{\rm fit } = p_{0}(1 + k_{1}\chi_{_{\rm BH}} +
	k_{2}\chi_{_{\rm BH}}^{2} + k_{3}\chi_{_{\rm BH}}^{3}) \\
	(1+ \ell_{1}\mathcal{C} + \ell_{2}\mathcal{C}^{2} +
	\ell_{1}\mathcal{C}^{3}) \,.
\end{align}
The above function provides a very good approximation of the separatrix
surface reported in Fig.~\ref{fig:3D_Omega_condition_separatrix} and the
corresponding best-fit results are given respectively by $p_{0} = 24.484$
$k_{1} = 1.464$, $k_{2} = 0.963$, $k_{3} = -0.162$, $\ell_{1} = -11.863$
$\ell_{2} = 50.877$, and $\ell_{3} = -77.050$. The analytic
representation of the critical surface provides a rather accurate
description of the numerical data with a coefficient of determination
$R^{2} = 0.988$, and relative differences that are smaller than $6.2\%$
for binaries with $Q\geq 1$.

\begin{figure}
  \centering
  \includegraphics[width=1.0\columnwidth]{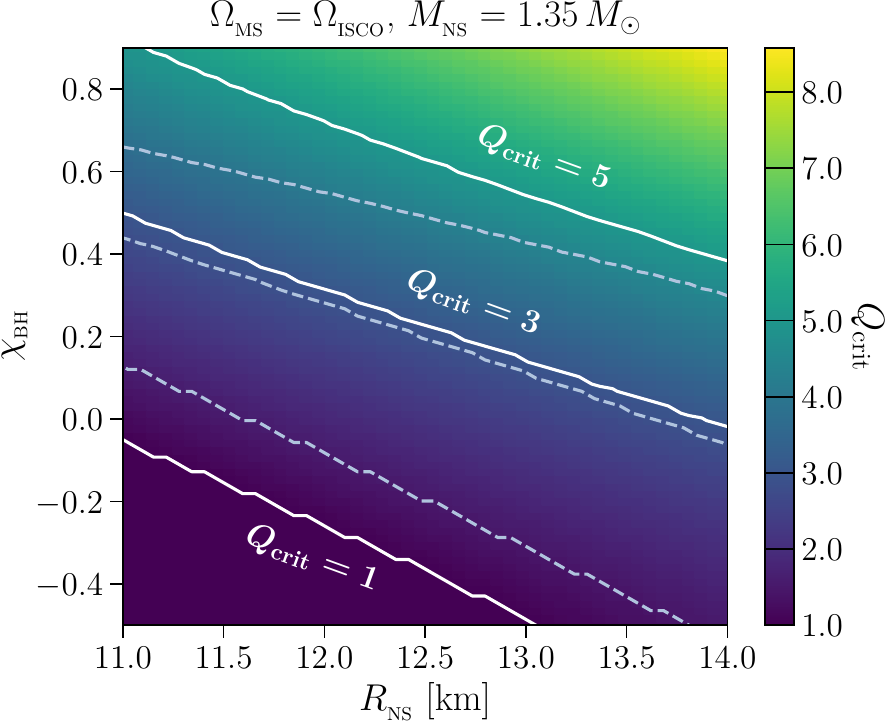}
  \caption{Critical value of the mass ratio along the separatric surface
    $\Omega_{_{\rm MS}}=\Omega_{_{\rm ISCO}}$ shown as a function of the
    BH spin $\chi_{_{\rm BH}}$ and NS radius for a fixed NS mass of
    $1.35\, M_{\odot}$. The expressions used are
    Eq.~\eqref{eq:Omega_MS_full_dependence} and
    Eq.~\eqref{eq:Omega_ISCO_full_dependence} with best-fit parameters
    for the intermediate EOS. White solid lines show contours at fixed
    $Q$, while dashed lines report the corresponding contours when
    considering a normalized remnant mass $\hat{M}_{\rm rem}\geq 10^{-2}$
    as estimated from the fitting model of Ref.~\cite{Foucart2018b}.}
  \label{fig:qcrit_F18_vs_T24_in_terms_of_RNS}
\end{figure}

Although Fig.~\ref{fig:3D_Omega_condition_separatrix} provides a complete
and comprehensive representation of the QE sequences, a different and
possibly more useful representation is also possible. In particular,
recalling that the behaviour of the separatrix between the mass-shedding
and plunge solutions is not very sensitive to the EOS and hence
``quasi-universal'' in the range of compactnesses $0.14 \lesssim
\mathcal{C} \lesssim 0.18$ (see
Fig.~\ref{fig:TD_vs_Plunge_AllEOS_and_F18} and discussion in
Sec.~\ref{sec:bhns_no_spin}), we can consider a NS with a given mass
$M_{_{\rm NS}}$ and map the dependence on the compactness into a
dependence on the stellar radius defined as $R_{_{\rm NS}} = M_{_{\rm
    NS}}/\mathcal{C}$. This is what is shown in
Fig.~\ref{fig:qcrit_F18_vs_T24_in_terms_of_RNS}, which reports the mass
ratio as a function of stellar radius and BH spin along the separatrix
surface $\Omega_{_{\rm MS}} = \Omega_{_{\rm ISCO}}$ and for a fixed NS
mass of $M_{_{\rm NS}} = 1.35\, M_{\odot}$. With this representation, and
after fixing a critical mass ratio $Q_{\rm crit}$, it is simple to
appreciate the radius above which the star will be tidally
disrupted. Furthermore, this mapping facilitates the comparison with the
similar Fig.~$2$ in Ref.~\cite{Foucart2020a}, which is built considering
a separatrix yielding the zero value for the remnant mass $\hat{M}_{\rm
  rem}$. To facilitate the comparison between the two approaches, and in
light of our finding in Fig.~\ref{fig:TD_vs_Plunge_AllEOS_and_F18}, we
report with gray dashed lines in
Fig.~\ref{fig:qcrit_F18_vs_T24_in_terms_of_RNS} the corresponding $Q_{\rm
  crit}$ contours obtained in terms of the remnant mass $\hat{M}_{\rm
  rem}=10^{-2}$. Overall, the contour lines are very similar despite the
systematic differences in the way they are computed (quasi-stationary
sequences or collection of dynamical simulations) and highlight that a
reference NS with a mass of $1.35\,M_{\odot}$ with a representative
radius of $\simeq 12.5\,{\rm km}$~\cite{Altiparmak:2022} and a mass ratio
$Q_{\rm crit}=3$ would only be disrupted before merger by a BH with spin
$\chi_{_{\rm BH}}\gtrsim 0.4$; conversely, the same star in a binary with
mass ratio of $Q_{\rm crit}=5$ would then require the BH to have a larger
spin of $\chi_{_{\rm BH}} \gtrsim 0.8$. Furthermore, in the case of a
GW230529-like event with $Q \sim 2.56$ ~\cite{LIGOScientific2024}, the
minimum BH spin would need to be $\chi_{_{\rm BH}}\gtrsim 0$ for tidal
disruption to occur.

\section{Conclusions}
\label{sec:summary}

In this first paper in a series about the equilibrium and dynamical
properties of BHNS binaries, we have investigated QE sequences of BHNS
initial data employing an improved version of the initial-data solver
\texttt{FUKA}, thus making use of what is arguably the largest collection
of individual BHNS configurations reported in the literature. In
addition, besides considering standard quantities for monitoring the
occurrence of mass-shedding or the occurrence of an unstable circular
orbit, we have also presented a novel approach based on the study of
tidal forces in terms of projections of the curvature tensor in a
suitable frame that provides a considerable amount of insight on the
forces and deformations taking place within the NS matter.

More specifically, we have first reported an extensive testing of the new
code by comparing its solutions for a $\Gamma=2$ polytropic EOS and
finding very good agreement with the results published in
Ref.~\cite{Taniguchi:2008a}. Having established the ability of the new
code to produce accurate initial data of BHNS binaries, we have used it
to investigate QE sequences of irrotational binaries, \ie where the BH is
nonrotating, under a variety of mass ratios and employing three distinct
EOSs extracted from an ensemble generated by uniform sampling with the
speed-of-sound parametrization and selected so as to provide a spread in
the EOS stiffness. In this way, we were able to determine and express
analytically not only the mass-shedding frequency as a function of the
mass ratio and stellar compactness $\Omega_{_{\rm MS}}(Q, \mathcal{C})$,
but also the related ISCO frequency $\Omega_{_{\rm ISCO}}(Q,
\mathcal{C})$.

The ability to express analytically $\Omega_{_{\rm MS}}$ and
$\Omega_{_{\rm ISCO}}$ has further allowed us to determine in the $(Q,
\mathcal{C})$ space the location of the separatrix for which
$\Omega_{_{\rm MS}}= \Omega_{_{\rm ISCO}}$ and thus to determine which
combinations of mass ratios and compactnesses in the BHNS binary lead to
either a ``mass-shedding'' scenario (\ie the tidal disruption of the NS
and the production of a massive merger remnant) or a ``plunge'' (\ie the
accretion of the NS without disruption). Interestingly, in a small but
nonzero range of compactnesses, \ie for $0.14 \lesssim \mathcal{C}
\lesssim 0.18$, the threshold between the two scenarios is only weakly
dependent on the EOS. In turn, this allows to draw conclusions on the
disruption scenario that can be expressed in terms of stellar radii.

The study of QE sequences of irrotational BHNS binaries has been further
extended by considering binaries in which the BH is spinning at different
rates and with spins that are either aligned or anti-aligned with the
orbital angular momentum. However, to make this study tractable from a
computational point of view, the investigation of spinning BHNS binaries
has been restricted to a single realistic EOS, namely the one with
intermediate stiffness. Also in this case, by determining the
mass-shedding and the ISCO frequencies and expressing them analytically
in the three-dimensional space of parameters $(Q, \mathcal{C},
\chi_{_{\rm BH}})$ it was possible to establish the location of the
separatrix $\Omega_{_{\rm MS}} = \Omega_{_{\rm ISCO}}$ and hence the
combinations of mass ratio, compactness and BH spin that can lead to a
tidal disruption and hence to a massive merger remnant. To the best of
our knowledge, this is the first time that the dependence of these
frequencies on the BH spin is investigated and that such a threshold
surface has been presented. Finally, exploiting the fact that for $0.14
\lesssim \mathcal{C} \lesssim 0.18$ the threshold between the two
scenarios is essentially EOS independent, we have mapped the predictions
about the disruption from the $(Q, \mathcal{C}, \chi_{_{\rm BH}})$ space
into a prediction in the $(Q, R_{_{\rm NS}}, \chi_{_{\rm BH}})$ space for
the reference case of a NS with a mass of $1.35\,M_{\odot}$. In turn,
this has revealed that, for instance, a reference value of
$\mathcal{C}=0.15$ can lead to disruption at $\chi_{_{\rm BH}}=0.8$ for
$Q=7$, while only at most $Q\approx 3$ is admissible for $\chi_{_{\rm
    BH}}=0.0$.

The analysis performed here can be extended in several directions. First,
our results could be compared with those obtained when abandoning the
assumption of conformal flatness, \ie where in the NS region,
non-diagonal terms of the spatial metric are nonzero. Such a comparison
would be especially useful to study the impact that conformal flatness
has on the tidal-force analysis presented here and on the influence that
BH spin exerts on the $\kappa$ and $E_{\rm b}$ diagnostics.
Unfortunately, we are not aware of initial-data solvers that can explore
this avenue at present. Second, it would be interesting and actually
important to consider additional EOSs to establish on more general
grounds whether a quasi-universal behaviour is present in the separatrix
surface between a tidal disruption and a plunge scenario. Third, given
that our QE sequences with an irrotational or aligned BH do not feature a
minimum in the $E_{\rm b}$ curves, it would be interesting to explore
other approaches that could potentially feature these minima, or be
capable of approaching closer separations, thus yielding better estimates
of the ISCO frequencies. Finally, a comparison could be carried out
between the mass-shedding and ISCO frequency estimates made from our QE
sequences and those instead inferred from fully dynamical
general-relativistic hydrodynamic simulations. Such a comparison for a
limited set of binaries will be presented in our accompanying paper
II~\cite{Topolski2024c}.

\section*{Data Availability}

The QE sequences data can be shared after a reasonable request.

\begin{acknowledgments}
We thank C. Ecker for sharing the ensemble of constraint-satisfying
equations of state and their PDF that we used for this work, as well as
for useful discussions and comments. KT expresses his gratitude to
R. Duqu\'e for extensive discussions on BHNS binaries. Support in funding
comes from the State of Hesse within the Research Cluster ELEMENTS
(Project ID 500/10.006), from the ERC Advanced Grant ``JETSET: Launching,
propagation and emission of relativistic jets from binary mergers and
across mass scales'' (Grant No. 884631). LR acknowledges the Walter
Greiner Gesellschaft zur Förderung der physikalischen Grundlagenforschung
e.V. through the Carl W. Fueck Laureatus Chair. The computations were
performed on HPE Apollo HAWK at the High Performance Computing Center
Stuttgart (HLRS) under the grants BNSMIC and BBHDISKS. ST gratefully
acknowledges support from NASA award ATP-80NSSC22K1898.
\end{acknowledgments}

\bibliographystyle{apsrev4-2}
%

\appendix

\section{3+1 Reconstruction of the spacetime Riemann curvature tensor}
\label{sec:3plus1_reconstruction_of_riemann}

This Appendix is dedicated to the presentation of additional details on
the diagnostic quantity that can be constructed by looking at the
magnitude of the Lie derivative along the helical Killing vector of the
extrinsic curvature tensor, as well as to a general exposition
of the necessary projections of the Riemann tensor used for computation
of tidal forces. We start by recalling that the XCTS system of
equations under the assumption of conformal flatness and in a 3+1
spacetime split provides the solutions for the spacetime variables
$(\psi, \alpha, \beta^{i})$ and for the matter variables $(E, p_{i},
S_{ij})$. To reach our goal we need to reconstruct the Riemann curvature
tensor from its projections and the right-hand side of the Einstein
equations. In our notation, we will mark with a top-left index ``4'' the
4-dimensional spacetime tensors associated with the full spacetime metric
and reserve the index $0$ to denote on which position the tensor has been
contracted with the normal vector to the hypersurface $n^{\mu}$.

Working in the conformally flat approximation $\gamma_{ij}=\psi^4
\tilde{\gamma}_{ij}$, the Ricci tensor $\tilde{R}_{ij}$ associated with
the background metric is zero and the spatial Ricci tensor can be
expressed fully in terms of the derivatives of the conformal factor
$\psi$ [see, \eg Eq.~(7.42) in~\cite{Gourgoulhon2012}]
\begin{align}
R_{ij} =& - 2 \tilde{D}_i\tilde{D}_j \ln \psi - 2 \tilde{D}_k \tilde{D}^k
\ln \psi \tilde{\gamma}_{ij} + \\
&4 \tilde{D}_i \ln \psi \tilde{D}_j \ln
\psi - 4 \tilde{D}_k \ln \psi \tilde{D}^k \ln \psi \tilde{\gamma}_{ij}\,.
\nonumber
\end{align}
Furthermore, because in 3 dimensions the Weyl tensor vanishes
identically, the spatial part of the Riemann tensor (note the lack of the
top-left index) is reconstructed by means of the Ricci decomposition from
the spatial Ricci tensor, spatial metric, and spatial Ricci scalar
$R=R_{ij}\gamma^{ij}$
\begin{align}
R_{ijkl} =&-R_{il} \gamma_{jk} + R_{jl} \gamma_{ik} + R_{ik} \gamma_{jl}
- R_{jk} \gamma_{il} +\\ &-\frac{1}{2} R (\gamma_{ik} \gamma_{jl} -
\gamma_{il}\gamma_{jk}) \,. \nonumber
\end{align}

The full projection onto the hypersurface can then be written using the
above tensor and the extrinsic curvature $K_{ij}$ by virtue of the
``Gauss relation'' [see, \eg Eq.~(3.73) in~\cite{Gourgoulhon2012}]
\begin{align}
^{4}R_{ijkl} = R_{ijkl} + K_{ik} K_{jl} - K_{il} K_{kj}\,.
\end{align}

The projection that involves one contraction with the normal vector is
given by the ``Codazzi-Mainardi relation'' [see, \eg Eq.~(3.81)
  in~\cite{Gourgoulhon2012}]
\begin{align}
^{4}R_{i0jk} = D_k K_{ji} - D_j K_{ki}\,,
\end{align}
while the twice-normal projection usually involves the so-called Ricci
equation, including the Lie derivative of the extrinsic curvature along
the normal direction. Here, we will instead use an expression that
requires the right-hand side of the trace-reversed Einstein equations,
\ie the contracted Gauss relation
\begin{align}
^{4}R_{i00j} \!=\! -R_{ij} - K K_{ij} + K^{l}_{~i} K_{lj}
  - 4 \pi [(S\! -\! E) \gamma_{ij} \!-\! 2S_{ij}] \,,
\end{align}
where we explicitly keep the trace of the extrinsic curvature even though
it is identically zero due to the maximal slicing condition that we employ for our
initial data. As part of the validation of our code, we evaluated the
Kretschmann curvature invariant $\mathcal{K} :=
R_{\alpha\beta\gamma\delta} R^{\alpha\beta\gamma\delta}$,
which can be obtained from the expression
\begin{align}
\mathcal{K} = {}^{4}R_{ijkl}{}^{4}R^{ijkl} - 4\;
        {}^{4}R_{i0jk}{}^{4}R^{i0jk} + 4\;
        {}^{4}R_{i00j}{}^{4}R^{i00j}\,,
\end{align}
where the raising of the spatial indices is performed with the spatial
metric $\gamma_{ij}$. By evaluating the Kretschmann scalar on a number of
analytical spacetimes that provide corresponding expressions in closed
form (including the Schwarzschild spacetime in isotropic coordinates, as
well as the de-Sitter metric), we have checked that the recovery of the
full Riemann tensor from 3+1 quantities is performed correctly.

To obtain the diagnostic field $\mathcal{L}_{\boldsymbol{\xi}} K_{ij}$
that we have reported in Fig.~\ref{fig:BHNS_XCTS_constraints_set}, we
make use of evolution equation for the extrinsic curvature $K_{ij}$ in
the $3+1$ split
\begin{align}
  \label{eq:dtKij_eq}
  \mathcal{L}_{\boldsymbol{\partial_{t}}}K_{ij} =&
  \mathcal{L}_{\boldsymbol{\beta}}K_{ij} - D_{i}D_{j}\alpha \\ &+
  \alpha\big{[} R_{ij} + K K_{ij} - 2K_{ik}K_{lj}\gamma^{kl} \big{]}
  \nonumber \\ &+4\pi \alpha\big{[} (S-E)\gamma_{ij} - 2 S_{ij}
    \big{]}\,, \nonumber
\end{align}
where the evolution is carried out along some coordinate time-like
four-vector $\boldsymbol{\partial_{t}}$, the covariant derivative
compatible with the spatial metric $\gamma_{ij}$ is denoted by $D_i$, and
the Lie derivative along the shift vector $\boldsymbol{\beta}$ is given
by $\mathcal{L}_{\boldsymbol{\beta}}$.

In the main text, we use Eq.~\eqref{eq:dtKij_eq} to compute the quantity
$\mathcal{L}_{\boldsymbol{\xi}}K_{xx}$, which measures the deviation of
the hypersurface from satisfying helical symmetry, where
$\boldsymbol{\partial_{t}} \rightarrow \boldsymbol{\xi}$ and
$\boldsymbol{\beta} \rightarrow \boldsymbol{B}$. The result is monitored
and reported in Table~\ref{tab:QEsequence_q5_interEOS} for various BHNS
binaries, and shown in Fig.~\ref{fig:BHNS_XCTS_constraints_set} for a
representative case. We also note that we compute the second covariant
derivative appearing in Eq.~\eqref{eq:dtKij_eq} as
\begin{align}
  D_{i} D_{j} \alpha = \tilde{D}_{i}\tilde{D}_{j} \alpha -
  \tilde{\Gamma}^{l}_{ij}\tilde{D}_{l}\alpha\,,
  \label{eq:ddalpha}
\end{align}
and that the Lie derivative along the shift vector follows after using the identity
\begin{align}
  \mathcal{L}_{\boldsymbol{\beta}}K_{ij}= \beta^{k} D_{k} K_{ij} + K_{ik}
  D_{j} \beta^{k} + K_{kj} D_{i} \beta^{k}\,,
  \label{eq:LieDer}
\end{align}
where the covariant derivative can be replaced by a partial derivative
because of the cancellation of the Christoffel symbols in a torsion-free
spacetime.

To complete the discussion of deviations from QE and quantify the impact
of BH spin and separation, in Table~\ref{tab:QEsequence_q5_interEOS} we
report volume-averaged measurements of
$\mathcal{L}_{\boldsymbol{\xi}}\ln \psi$ and
$\mathcal{L}_{\boldsymbol{\xi}}K_{ij}$, calculated as follows. We define integrals
over a volume $V$
\begin{align}
	\Vert \mathcal{L}_{\boldsymbol{\xi}}\ln \psi \Vert
		\coloneqq \frac{1}{\mathcal{V}} \int_{V} \vert
		\mathcal{L}_{\boldsymbol{\xi}}\ln \psi \vert
		\sqrt{\gamma}d^{3} x
\end{align}
and
\begin{align}
	\Vert \mathcal{L}_{\boldsymbol{\xi}}K_{ij} \Vert
		\coloneqq \frac{1}{\mathcal{V}} \int_{V} \big{(}
		\mathcal{L}_{\boldsymbol{\xi}}K_{ij}
		\mathcal{L}_{\boldsymbol{\xi}}K^{ij} \big{)}^{1/2}
		\sqrt{\gamma}d^{3} x
\end{align}
where $\mathcal{V}=\int_{V}\sqrt{\gamma}d^{3}x$ is the volume of a
spherical region serving as the normalization factor, and $\gamma$ the
determinant of the spatial metric $\gamma_{ij}$. The integrals are
evaluated numerically using the Monte-Carlo method with around $n=6000$
points within a sphere of radius $r=60M_{\odot}$, measured from the
origin of the coordinate system. The black hole's apparent horizon is
excluded from the computation domain. An inspection of the last two
columns of Table~\ref{tab:QEsequence_q5_interEOS} supports the
observation made already in reference to
Fig.~\ref{fig:BHNS_XCTS_constraints_set}, namely that
$\mathcal{L}_{\boldsymbol{\xi}}\ln \psi$ is on average an order of
magnitude smaller than $\mathcal{L}_{\boldsymbol{\xi}}K_{ij}$.

\section{Dependence of $E_{\rm b}$ on $\chi_{_{\rm BH}}$}
\label{sec:Eb_spin_dependence}

In Sec.~\ref{sec:bhns_no_spin} of the main text we have presented the
behaviour of the binding energy as a function of the binary orbital
angular frequency and we have presented in
Fig.~\ref{fig:bhns_seq_with_spin_qhat5} a summary of the results obtained
when considering BHs with different spins. We here go back to that figure
and discuss the nontrivial dependence of the binding energy on the BH
spin in the case of aligned spins and that, to the best of our knowledge,
has not been discussed before in the literature.

\begin{figure}
  \centering
  \includegraphics[width=0.98\columnwidth]{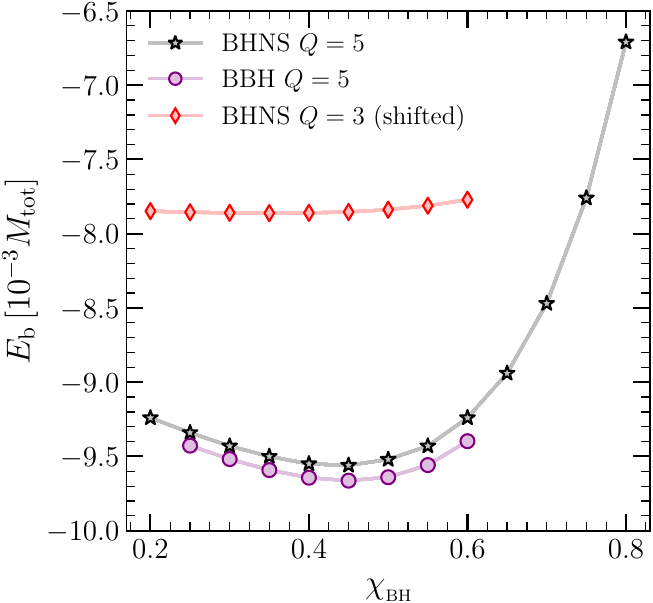}
  \hskip 0.5cm
  \caption{Dependence of the binding energy on the BH spin for
    binaries at a fixed separation of $d=40\, M_{\odot}$. Black stars and
    red diamonds refer to BHNS binaries with $Q=5$ and $Q=3$,
    respectively (the latter have been rescaled by a fixed amount to be
    shown on the same plot). Also shown with purple circles are BBH
    binaries with equivalent constituent masses and spins.}
  \label{fig:binding_energy_spin}
\end{figure}

More specifically, let us consider as representative examples the QE
sequences with constant mass ratio shown in
Fig.~\ref{fig:bhns_seq_with_spin_qhat5} and examine the variation of the
binding energy in the case of aligned (\ie positive) BH spins. This is
reported in Fig.~\ref{fig:binding_energy_spin}, where we show the values
of binding energy at a fixed separation of $d = 40\, M_{\odot}$ for the
sequences with $Q = 3, 5$ (black and red symbols and lines) and having
total masses $M_{\rm tot}=5.200\, M_{\odot}$ and $M_{\rm tot} = 7.800\,
M_{\odot}$, respectively (this translates to $d/M_{\rm tot} \approx 7.7$
and $d / M_{\rm tot} \approx 5.13$ in the two cases). Clearly, the
binding energy does not have a monotonic behaviour with $\chi_{_{\rm
    BH}}$, but first decreases with spin and then increases, with a local
minimum at around $\chi_{_{\rm BH}} \approx 0.45$. Note that this
behaviour is present in both BHNS sequences (the sequence with $Q = 3$ is
shifted by a quantity $-6 \times 10^{-3}\,M_{\rm tot}$, with $M_{\rm
  tot}=5.200\, M_{\odot}$ to make it visible on the same plot; the
minimum of $E_{\rm b}$ moves to around $\chi_{_{\rm BH}} \approx 0.35$)
but is not a unique feature of BHNS QE binaries. Indeed, this is show
eloquently by reporting in Fig.~\ref{fig:binding_energy_spin} also QE
sequences of BBHs having the same mass ratio of $Q = 5$ (purple symbols
and line). Bearing in mind that BHs have zero tidal deformability and
hence the corresponding binding energy at the same orbital angular
velocity is (slightly) smaller, the BBH sequence exhibits essentially
the same behaviour as the BHNS sequence. Of course, the fact that BBHs
behave in a similar manner is helpful, but does not necessarily help
explain the origin of this behaviour.

Hence, in order to gain a better understanding of the underlying physical
grounds, we have extended the investigation presented in
Fig.~\ref{fig:binding_energy_spin}. More specifically, since the QE
sequences considered refer to rather close binaries, \ie with $d / M_{\rm
  tot} \approx 6$, for which the assumptions behind the calculation of
the initial data may be violated (\eg conformal flatness introducing
non-physical radiation content, quasi-circularity) we have explored the
behaviour of $E_{\rm b}(\chi_{_{\rm BH}})$ also for binaries that are
more widely separated, \ie with $d / M_{\rm tot} \approx 12$. In this
way, we have found that the non-monotonic behaviour of $E_{\rm
  b}(\chi_{_{\rm BH}})$ is still present, although to a smaller degree,
hence excluding the above-mentioned violations as the sole source of the
behaviour.

\begin{figure}
  \centering
  \includegraphics[width=0.98\columnwidth]{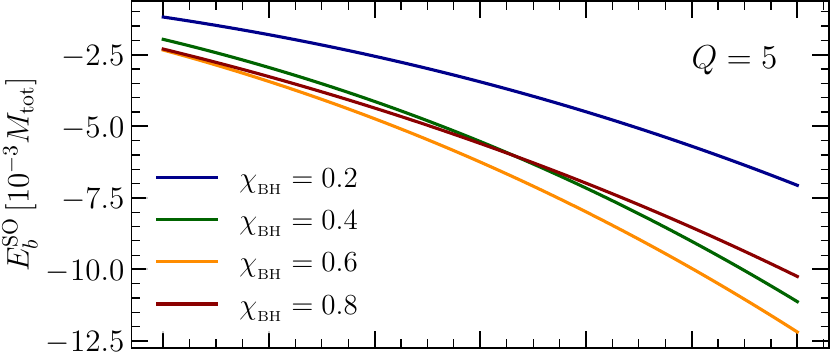}
  \includegraphics[width=0.98\columnwidth]{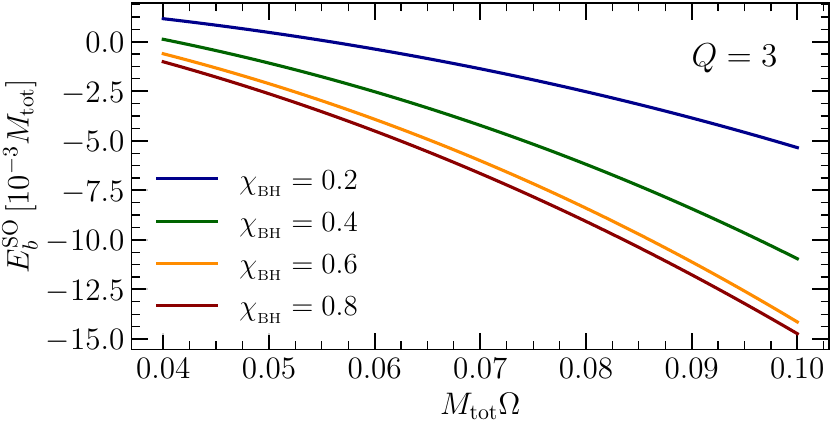}
  \caption{Spin-orbit contribution to the binding energy of a BBH system
    according to Eq.~(5.4) in Ref.~\cite{Levi2015a} and shown as a
    function of the spin of the primary. The data refers to mass ratios
    $Q=5$ (top panel) and $Q=3$ (bottom panel).}
	\label{fig:Eb_SO_expr_q3_q5}
\end{figure}

Next, we have considered the contribution to the binding energy related
to the spin that can be deduced from the post-Newtonian (PN)
approximation. In particular, it is possible to isolate the
spin-dependent contributions to the binding energy up to 3.5 PN order
including cubic terms in spin [see, \eg Eq.~(5.4) of
  Ref.~\cite{Levi2015a}] and evaluate them for spins in the range
$\chi_{_{\rm BH}}=0.2-0.8$, mass ratios $Q=3,\, 5$, and for orbital
frequencies $M_{\rm tot}\Omega \gtrsim 0.05$. For simplicity, the
expressions are all evaluated with so-called Wilson coefficients all set
to unity and hence corresponding to a BBH system\footnote{Wilson
coefficients arise as constants in the Lagrangian of the point-particle
effective action; for more details, see \eg~\cite{Levi2015b,
  Mandal2023}].} [see~\cite{Laarakkers1997, Pappas2012} in the case of a
    NS]. Our Fig.~\ref{fig:binding_energy_spin} suggests that
  qualitatively similar conclusions should be drawn also for a BHNS
  binary at these frequencies. The evaluated spin-orbit contributions to
  the binding energy, $E^{_{\rm SO}}_{\rm b}$, are presented in
  Fig.\ref{fig:Eb_SO_expr_q3_q5} and can constitute a significant part of
  the total binding energy if the primary is rapidly spinning. Clearly,
  while these contributions vary monotonically with BH spin for the mass
  ratio $Q=3$ (bottom panel), this is not the case for larger mass
  asymmetries of $Q=5$ (top panel), which is consistent with the QE
  picture presented in Fig.~\ref{fig:binding_energy_spin}. Interestingly,
  we find that the inclusion of cubic terms in spin is essential to
  observe the crossing of $\chi_{_{\rm BH}}=0.2$ and $\chi_{_{\rm
      BH}}=0.8$ curves at $M_{\rm tot}\Omega \approx 0.05$ in
  Fig.\ref{fig:Eb_SO_expr_q3_q5}, which otherwise shifts to $M_{\rm
    tot}\Omega \approx 0.075$. Overall, while we have carried out
  investigations in a number of directions, the fact that the binding
  energy does not behave monotonically remains unclear and hence a
  perfect opportunity for additional studies, possibly employing
  dynamical simulations, as done in Ref.~\cite{Ossokine2017}, or
  validation against Post-Newtonian predictions with the inclusion of
  spin as an extension of Ref.~\cite{Berti08}, to determine whether this
  behaviour is robust.

\section{Supplementary information and figures}
\label{sec:Omega_MS_ISCO_Soft_Stiff}

In this Appendix we provide additional information that supplements the
one illustrated in the main text. While this data is available also in
the publicly released data-set, we provide it here as well for
completeness. In addition, we present figures that are similar to those
already presented in the main text, but refer to the different QE
sequences of binaries explored in this work.

We start by reporting in Tab.~\ref{tab:QEsequence_q5_interEOS} the
complete set of data relative to the binaries discussed in this paper,
for a fixed mass ratio of $Q=5$ and with the EOS of intermediate
stiffness. These diagnostics can be readily obtained for all the QE
sequences in this work by inspecting the released publicly available
data.

\begin{table*}[t]
  \renewcommand{\arraystretch}{1.2}
  \setlength{\tabcolsep}{5pt}
  \begin{tabular}{|c|c|c|c|c|c|c|c|c|c|c}
      \hline
      $d$ & $d/M_{\rm tot}$ & $M_{\rm tot}\Omega $ & $\kappa$ & $E_{\rm b}/M_{\rm tot}$ & $M_{_{\rm ADM}}$ &  $J_{_{\rm ADM}}$ & $M_{_{\rm K}}$ & $\delta M$ &
	  $\Vert \mathcal{L}_{\boldsymbol{\xi}}\ln \psi \Vert$ & $\Vert \mathcal{L}_{\boldsymbol{\xi}}K_{ij} \Vert$
  \\
      $[M_{\odot}]$ & & $[\times 10^{-2}]$ & & $[\times 10^{-3}]$ &
      $[M_{\odot}]$ & $[M_{\odot}^{2}]$ & $[M_{\odot}]$ &  $[\times
	  10^{-4}]$ & $[\times 10^{-6}]$  & $[\times 10^{-5}]$\\
      \hline
      \multicolumn{11}{c}{ $\chi_{_{\rm BH}}=-0.8$  } \\
      \hline
      $60.000$ & $7.692$ & $4.070$ & $0.964$ & $-2.450$ & $7.78085$ & $-0.772$ & $7.78562$ & $6.130$ & $5.290$  & $3.403$ \\
      $55.000$ & $7.051$ & $4.587$ & $0.950$ & $-2.650$ & $7.77932$ & $-1.032$ & $7.78426$ & $6.350$ & $6.647$  & $4.048$ \\
      $50.000$ & $6.410$ & $5.225$ & $0.929$ & $-2.850$ & $7.77776$ & $-1.206$ & $7.78311$ & $6.879$ & $9.137$  & $5.788$ \\
      $45.000$ & $5.769$ & $6.029$ & $0.894$ & $-2.870$ & $7.77758$ & $-1.196$ & $7.78339$ & $7.470$ & $9.505$  & $5.941$ \\
      $42.500$ & $5.449$ & $6.512$ & $0.867$ & $-2.790$ & $7.77821$ & $-1.097$ & $7.78427$ & $7.791$ & $11.551$ & $6.892$ \\
      $40.000$ & $5.128$ & $7.065$ & $0.831$ & $-2.620$ & $7.77960$ & $-0.905$ & $7.78607$ & $8.317$ & $12.427$ & $7.458$ \\
      $39.500$ & $5.064$ & $7.185$ & $0.823$ & $-2.560$ & $7.78002$ & $-0.853$ & $7.78657$ & $8.419$ & $12.829$ & $7.711$ \\
      $39.000$ & $5.000$ & $7.309$ & $0.813$ & $-2.500$ & $7.78050$ & $-0.794$ & $7.78716$ & $8.560$ & $13.281$ & $7.995$ \\
      \hline
      \multicolumn{11}{c}{ $\chi_{_{\rm BH}}=-0.4$  } \\
      \hline
      $60.000$ & $7.692$ & $4.002$ & $0.967$ & $-6.460$ & $7.74959$ & $14.595$ & $7.75008$ & $0.632$ & $1.777$ & $1.469$ \\
      $45.000$ & $5.769$ & $5.880$ & $0.908$ & $-7.390$ & $7.74237$ & $13.451$ & $7.74323$ & $1.111$ & $3.606$ & $2.645$ \\
      $42.500$ & $5.449$ & $6.338$ & $0.887$ & $-7.490$ & $7.74161$ & $13.360$ & $7.74254$ & $1.201$ & $4.168$ & $2.945$ \\
      $40.000$ & $5.128$ & $6.860$ & $0.859$ & $-7.510$ & $7.74143$ & $13.329$ & $7.74242$ & $1.279$ & $4.915$ & $3.529$ \\
      $39.000$ & $5.000$ & $7.090$ & $0.845$ & $-7.500$ & $7.74148$ & $13.336$ & $7.74251$ & $1.330$ & $5.350$ & $3.828$ \\
      $38.500$ & $4.936$ & $7.209$ & $0.837$ & $-7.490$ & $7.74158$ & $13.346$ & $7.74262$ & $1.343$ & $5.599$ & $3.977$ \\
      \hline
      \multicolumn{11}{c}{ $\chi_{_{\rm BH}}=0$  } \\
      \hline
      $60.000$ & $7.692$ & $3.946$ & $0.970$ & $-6.940$ & $7.74589$ & $30.212$ & $7.74606$ & $0.219$ & $0.766$ & $0.838$  \\
      $55.000$ & $7.051$ & $4.431$ & $0.960$ & $-7.340$ & $7.74278$ & $29.586$ & $7.74294$ & $0.207$ & $0.992$ & $1.057$  \\
      $50.000$ & $6.410$ & $5.021$ & $0.944$ & $-7.820$ & $7.73902$ & $29.016$ & $7.73926$ & $0.310$ & $1.314$ & $1.358$  \\
      $45.000$ & $5.769$ & $5.758$ & $0.919$ & $-8.280$ & $7.73540$ & $28.491$ & $7.73575$ & $0.452$ & $1.800$ & $1.785$  \\
      $42.500$ & $5.449$ & $6.197$ & $0.902$ & $-8.510$ & $7.73358$ & $28.253$ & $7.73396$ & $0.491$ & $2.128$ & $2.075$  \\
      $40.000$ & $5.128$ & $6.693$ & $0.878$ & $-8.730$ & $7.73194$ & $28.042$ & $7.73228$ & $0.440$ & $2.562$ & $2.423$  \\
      $39.500$ & $5.064$ & $6.801$ & $0.873$ & $-8.760$ & $7.73164$ & $28.005$ & $7.73197$ & $0.427$ & $2.664$ & $2.502$  \\
      $39.000$ & $5.000$ & $6.911$ & $0.867$ & $-8.810$ & $7.73132$ & $27.969$ & $7.73165$ & $0.427$ & $2.770$ & $2.586$  \\
      $38.500$ & $4.936$ & $7.024$ & $0.860$ & $-8.840$ & $7.73104$ & $27.934$ & $7.73137$ & $0.427$ & $2.884$ & $2.674$  \\
      $38.250$ & $4.904$ & $7.082$ & $0.857$ & $-8.860$ & $7.73092$ & $27.919$ & $7.73125$ & $0.427$ & $2.940$ & $2.720$  \\
      \hline
      \multicolumn{11}{c}{ $\chi_{_{\rm BH}}=0.4$  } \\
      \hline
      $60.000$ & $7.692$ & $3.897$ & $0.972$ & $-7.110$ & $7.74456$ & $45.909$ & $7.74488$ & $0.413$ & $1.499$ & $1.320$ \\
      $55.000$ & $7.051$ & $4.370$ & $0.963$ & $-7.590$ & $7.74082$ & $45.183$ & $7.74114$ & $0.413$ & $1.767$ & $1.401$ \\
      $45.000$ & $5.769$ & $5.655$ & $0.928$ & $-8.810$ & $7.73130$ & $43.705$ & $7.73178$ & $0.621$ & $2.677$ & $2.218$ \\
      $42.500$ & $5.449$ & $6.078$ & $0.912$ & $-9.150$ & $7.72862$ & $43.345$ & $7.72912$ & $0.647$ & $3.016$ & $2.432$ \\
      $40.000$ & $5.128$ & $6.555$ & $0.893$ & $-9.520$ & $7.72577$ & $42.990$ & $7.72626$ & $0.634$ & $3.412$ & $2.854$ \\
      $39.500$ & $5.064$ & $6.657$ & $0.888$ & $-9.590$ & $7.72522$ & $42.920$ & $7.72570$ & $0.621$ & $3.496$ & $2.923$ \\
      $39.000$ & $5.000$ & $6.763$ & $0.883$ & $-9.660$ & $7.72468$ & $42.852$ & $7.72515$ & $0.608$ & $3.590$ & $3.079$ \\
      $38.500$ & $4.936$ & $6.872$ & $0.878$ & $-9.730$ & $7.72413$ & $42.785$ & $7.72459$ & $0.596$ & $3.700$ & $3.151$ \\
      $38.000$ & $4.872$ & $6.982$ & $0.873$ & $-9.860$ & $7.72313$ & $42.706$ & $7.72343$ & $0.388$ & $3.811$ & $3.162$ \\
      $37.500$ & $4.808$ & $7.097$ & $0.868$ & $-9.940$ & $7.72251$ & $42.638$ & $7.72280$ & $0.376$ & $3.926$ & $3.235$ \\
      \hline
      \multicolumn{11}{c}{ $\chi_{_{\rm BH}}=0.8$  } \\
      \hline
      $60.000$ & $7.692$ & $3.858$ & $0.974$ & $-3.750$ & $7.77072$ & $61.816$ & $7.77496$ & $5.456$ & $4.874$ & $3.041$ \\
      $45.000$ & $5.769$ & $5.572$ & $0.933$ & $-5.770$ & $7.75502$ & $59.183$ & $7.75948$ & $5.751$ & $7.444$ & $4.838$ \\
      $42.500$ & $5.449$ & $5.982$ & $0.920$ & $-6.210$ & $7.75160$ & $58.720$ & $7.75612$ & $5.831$ & $7.968$ & $5.600$ \\
      $40.000$ & $5.128$ & $6.443$ & $0.903$ & $-6.680$ & $7.74786$ & $58.248$ & $7.75241$ & $5.873$ & $8.520$ & $5.973$ \\
      $39.500$ & $5.064$ & $6.542$ & $0.899$ & $-6.790$ & $7.74706$ & $58.152$ & $7.75161$ & $5.873$ & $8.779$ & $6.099$ \\
      $39.000$ & $5.000$ & $6.644$ & $0.895$ & $-6.920$ & $7.74602$ & $58.051$ & $7.75045$ & $5.719$ & $9.050$ & $6.245$ \\
      $38.500$ & $4.936$ & $6.749$ & $0.890$ & $-6.980$ & $7.74552$ & $57.961$ & $7.75006$ & $5.861$ & $9.321$ & $6.406$ \\
      \hline
    \end{tabular}
    \caption{Most relevant quantities for the QE sequences presented in
      this paper and relative to the EOS with intermediate stiffness and
      binaries with mass ratio $Q=5$, NS mass of $M_{\rm NS}=1.3 \,
      M_{\odot}$, total mass $M_{\rm tot}= 7.8\, M_{\odot}$ (see the
      Data-Availability section for information on the full
      data-set). For each sequence with constant BH spin $\chi_{_{\rm
          BH}}$, we report: the initial separation $d_{0}$, the
      dimensionless initial separation $d_{0}/M_{\rm tot}$, the
      dimensionless orbital angular velocity $M_{\rm tot}\Omega$, the
      mass-shedding diagnostic $\kappa$, the dimensionless binding energy
      $E_{\rm b}/M_{\rm tot}$, the total ADM mass $M_{_{\rm ADM}}$ and
      ADM momentum $J_{_{\rm ADM}}$, the Komar mass $M_{_{\rm K}}$, the
      (dimensionless) virial error $\delta M$ [\cf
        Eq.~\eqref{eq:virial_error}], as well as volume-averaged
      deviations from QE $\Vert \mathcal{L}_{\boldsymbol{\xi}}\ln \psi
      \Vert$ and $\Vert \mathcal{L}_{\boldsymbol{\xi}}K_{ij} \Vert$. }
\label{tab:QEsequence_q5_interEOS}
\end{table*}

\begin{figure*}
  \centering
  \includegraphics[width=0.49\columnwidth]{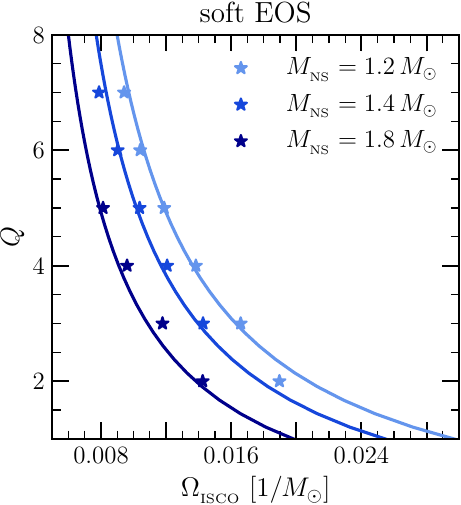}
  \hskip 0.25cm
  \includegraphics[width=0.49\columnwidth]{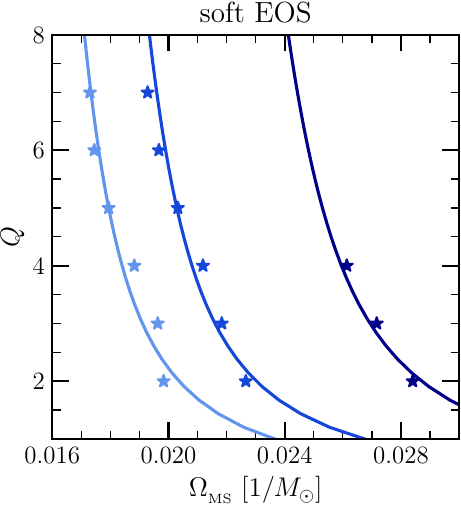}
  \hskip 0.25cm
  \includegraphics[width=0.49\columnwidth]{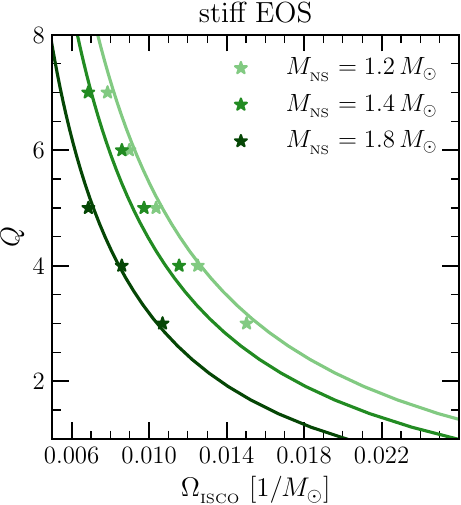}
  \hskip 0.25cm
  \includegraphics[width=0.49\columnwidth]{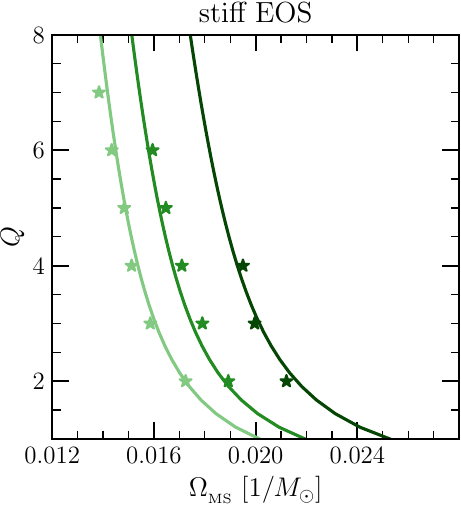}
  \caption{(The same as Fig.~\ref{fig:Intermediate_Omegas}, but
    for the soft EOS (leftmost panels) and for the stiff EOS (rightmost panels).}
  \label{fig:Soft_Stiff_Omegas}
\end{figure*}

Next, we present in Fig.~\ref{fig:Soft_Stiff_Omegas} a representation of
the $\Omega_{_{\rm MS}}$ and $\Omega_{_{\rm ISCO}}$ frequencies analogous
to the one shown in Fig.~\ref{fig:Intermediate_Omegas}, but for the soft
(two left-most panels) and stiff EOSs (two right-most panels). The ADM
masses of the NSs stars are fixed to the same values $1.2,1,4,1.8\,
M_{\odot}$, alongside the fits with the functional forms
Eq.~\eqref{eq:mtotomega_fits_ms} and \eqref{eq:mtotomega_fits_isco} with
corresponding best-fit parameters reported in
Tab.~\ref{tab:fitting_params_mtotomega} with different colours.

For the $\Omega_{_{\rm MS}}$ frequencies, the impact of different
compactness at the same ADM mass is clearly visible. For a given NS mass
$M_{_{\rm NS}}$, the $\Omega_{_{\rm MS}}$ shifts to higher frequencies
for the soft EOS, indicating a later onset of mass shedding. On the other
hand, the stiff EOS leads to the NS disrupting more easily than the
intermediate and soft ones, and the values of $\Omega_{_{\rm MS}}$ are
shifted to lower frequencies accordingly. Noteworthy, for the soft EOS,
the separation between the $1.4 \, M_{\odot}$ and $1.8\, M_{\odot}$
sequences is larger than for any of the other EOSs. This is due to the
largest compactness featured in this paper for the $1.8\, M_{\odot}$ NS
and given by $\mathcal{C}=0.234$. On the other hand, with regard to the
$\Omega_{_{\rm ISCO}}$ frequencies, the two EOSs feature a similar range
and dependence on the mass ratio, which is comparable to the one for the
intermediate EOS discussed in the main text.

\begin{figure*}
  \centering
  \includegraphics[width=0.95\columnwidth]{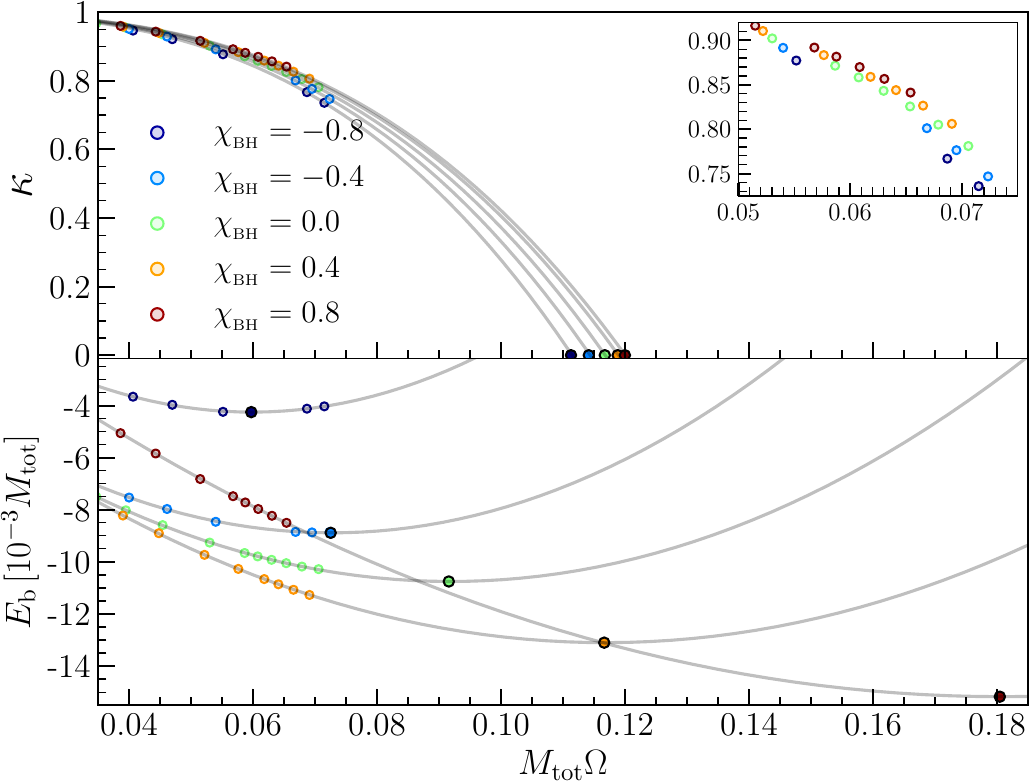}
  \hskip 0.5cm
  \includegraphics[width=0.95\columnwidth]{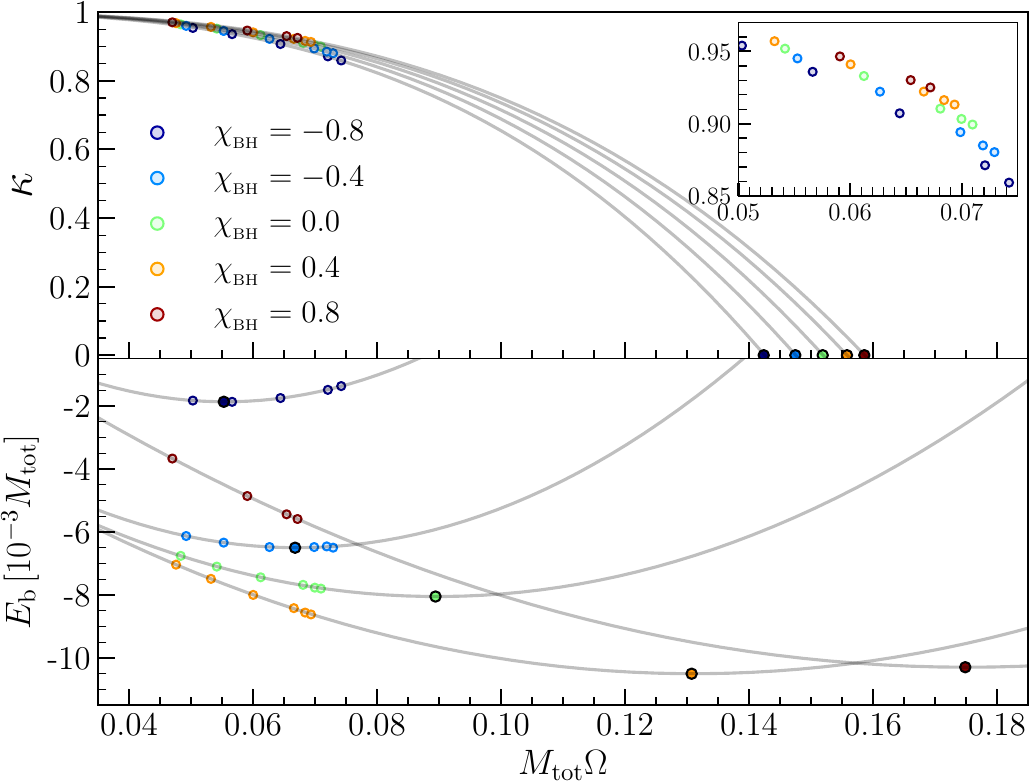}
  \caption{The same as in Fig.~\ref{fig:bhns_seq_with_spin_qhat5}, but
    for BHNS binaries having either mass ratio $Q=4$ (left panel) or
    $Q=6$ (right panel).}
  \label{fig:bhns_seq_with_spin_qhat4_6}
\end{figure*}

Finally, Fig.~\ref{fig:bhns_seq_with_spin_qhat4_6} features a
complementary picture of the QE sequences with a spinning BH, analogous
to Fig.~\ref{fig:bhns_seq_with_spin_qhat5} in the main text, but with
mass ratios $Q=4$ and $Q=6$. While the impact of mass ratio has already
been discussed at length in Sec.~\ref{sec:results}, the impact of BH spin
is largely independent of the mass ratio and thus the curves at different
$\chi_{_{\rm BH}}$ follow the same pattern. This justifies the rescaling
approach that we have chosen in Sec.~\ref{sec:results}; only a small
mixed term $\propto q\,\chi_{_{\rm BH}}$ is needed for the $\Omega_{_{\rm
    MS}}$ frequency, whereas the rescaled $\Omega_{_{\rm ISCO}}$ values
for different mass ratios lie along a curve in
Fig.~\ref{fig:ISCO_MS_spin_dependence} that is best described by a purely
$\chi_{_{\rm BH}}$ dependence.

\end{document}